\documentclass[10pt,final,doublecolumn]{IEEEtran}
\IEEEoverridecommandlockouts
\usepackage{amsmath}
\usepackage{amssymb}
\usepackage{amsthm}
\usepackage{mathrsfs}
\usepackage{latexsym}
\usepackage{graphicx}
\usepackage{bbding}
\usepackage{indentfirst}
\usepackage{cases}
\usepackage{supertabular}
\usepackage{algorithm,algorithmic}
\usepackage{subeqnarray}
\usepackage{color}
\usepackage{bm}
\usepackage{stfloats}
\usepackage{subfigure}
\usepackage{array}
\usepackage{algorithm,float}
\usepackage{multirow}

\IEEEoverridecommandlockouts
\allowdisplaybreaks[4]

\makeatletter
\newenvironment{breakablealgorithm}
{% \begin{breakablealgorithm}
		\begin{center}
			\refstepcounter{algorithm}% New algorithm
			\hrule height.8pt depth0pt \kern2pt% \@fs@pre for \@fs@ruled 画线
			\renewcommand{\caption}[2][\relax]{% Make a new \caption
				{\raggedright\textbf{\ALG@name~\thealgorithm} ##2\par}%
				\ifx\relax##1\relax % #1 is \relax
				\addcontentsline{loa}{algorithm}{\protect\numberline{\thealgorithm}##2}%
				\else % #1 is not \relax
				\addcontentsline{loa}{algorithm}{\protect\numberline{\thealgorithm}##1}%
				\fi
				\kern2pt\hrule\kern2pt
			}
		}{% \end{breakablealgorithm}
		\kern2pt\hrule\relax% \@fs@post for \@fs@ruled 画线
	\end{center}
}
\makeatother

\begin{document}
\title{QoS-Driven Satellite Constellation Design for LEO Satellite Internet of Things}
\author{\IEEEauthorblockN{
\normalsize Ming Ying, Xiaoming Chen, Qiao Qi, and Zhaoyang Zhang}
\thanks{Ming Ying, Xiaoming Chen, and Zhaoyang Zhang are with the College of Information Science and Electronic Engineering, Zhejiang University, Hangzhou 310027, China (e-mail:\{ming\_ying, chen\_xiaoming, ning\_ming\}@zju.edu.cn). Qiao Qi is with the School of Information Science and Technology, Hangzhou Normal University, Hangzhou 311121, China (email: qiqiao@hznu.edu.cn). }
}\maketitle

\begin{abstract}
    Low Earth orbit (LEO) satellite Internet of Things (IoT) has been identified as one of the important components of the sixth-generation (6G) non-terrestrial networks (NTN) to provide ubiquitous connectivity. Due to the low orbit altitude and high mobility, a massive number of satellites are required to form a global continuous coverage constellation, leading to a high construction cost. To this end, this paper proposes a LEO satellite IoT constellation design algorithm with the goal of minimizing the total cost while satisfying quality of service (QoS) requirements in terms of coverage ratio and communication quality. Specifically, with a novel fitness function and efficient algorithm's operators, the proposed algorithm converges more quickly and achieves lower constellation construction cost compared to baseline algorithms under the same QoS requirements. Theoretical analysis proves the global and fast convergence of the proposed algorithm due to a novel fitness function. Finally, extensive simulation results confirm the effectiveness of the proposed algorithm in LEO satellite IoT constellation design.
\end{abstract}

\providecommand{\keywords}[1]{\textbf{\textit{Index Terms---}} #1}

\section{Introduction}
    Nowadays, a massive number of IoT devices have been deployed all over the world for various applications, e.g., intelligent transport, smart healthcare, and factory automation, which have notably enhanced people's quality of life \cite{iot1}. Hence, International Telecommunication Union (ITU) has identified ubiquitous connectivity as one of the usage cases of 6G wireless networks  \cite{iot2}-\cite{iot4}. However, the {remote and isolated locations of some IoT devices} and unstable wireless communication environment limit the traditional terrestrial cellular networks. For instance, it is difficult to construct new base stations for IoT applications in mountains, oceans, deserts and other remote areas due to the great challenge for construction and the heavy cost for maintenance. In this context, the Direct-to-Satellite (DtS) communication scheme is utilized to establish massive access\footnote{According to \cite{iot5}, massive access refers to that a massive number of wireless devices access the system randomly. The tremendous growth in the number of connected devices requires a fundamental rethinking of the conventional multiple access technologies in favor of new schemes suited for massive random access. Specifically, massive access can potentially support the random access millions of devices in an area of one square kilometer.
    	} for IoT devices deployed in remote areas. Specifically, DtS does not need a terrestrial gateway or other relay stations to propagate the signals, but the devices connect directly to the LEO satellite  \cite{F1}- \cite{F3}. It can lead to reduced latency, improved data transfer speeds, and enhanced constellation coverage, especially in remote or geographically challenging areas \cite{F4}. Additionally, DtS communication can offer greater flexibility and mobility, making it a valuable option for applications such as maritime and aerial communication, where continuous connectivity is essential \cite{F5}. However, one LEO satellite moves very fast and thus cannot provide persistent coverage to a specific area. Therefore, it is necessary to construct a LEO satellite network to provide seamless coverage, sufficient capacity and {large-scale} connectivity for IoT devices.

    Generally, the construction of LEO satellite networks can be divided into several stages, in which one of the most important being satellite constellation design, as it directly affects the overall system performance. A LEO satellite constellation can be regarded as a number of satellites deployed into the pre-determined orbits to form a satellite network. To provide coverage for Earth, there have been several classical constellation architectures, such as polar orbit constellation \cite{polar} and Walker Delta constellation \cite{walker}. Both of these constellations are favored for their symmetry, which plays a crucial role in ensuring efficient coverage and system performance. Generally speaking, polar orbit constellation can provide excellent global coverage. As Earth rotates underneath the orbit, the satellites can cover the entire surface of Earth over time. {The Walker Delta constellation is also a specific type of satellite constellation designed to provide predictable coverage within a latitude-band region due to its symmetry.} For instance, the first stage of Starlink is constructed based on Walker Delta constellation \cite{starlink}. It has 24 orbit planes with 66 LEO satellites on each plane at 550 km altitude. The orbit inclination is 53$^\circ$ and the phase factor is 11. Moreover, Flower constellation is a special type of Walker delta constellation with only one satellite in each orbital plane, aimed to achieve high global coverage with a small number of satellites \cite{Flower1}, \cite{Flower2}. {In general, quality-of-service (QoS) requirements of terrestrial IoT devices and deployment costs of a large number of satellites are two key factors of satellite constellation design.} {Particularly, the QoS requirements of the satellite constellation often include coverage ratio and communication quality, while the costs of the satellite constellation usually consist of three parts: manufacturing cost, launching cost and insurance cost \cite{cost}, \cite{cost2}.} Therefore, the design of LEO satellite constellation should balance the QoS requirements of LEO satellite IoT and the costs of the LEO satellite deployment.

\begin{table*}
	\renewcommand{\arraystretch}{1.3}
	\centering
	\small
	\caption{Key Limitations of Existing Techniques and Proposed Solutions}\label{Simulation}
	\begin{tabular}{|c|c|c|c|c|c|c|}
		\hline
		\multirow{2}{*}{\textbf{References}}& \multirow{2}{*}{\textbf{Optimization Algorithms}} &\multirow{2}{*}{\textbf{Devices' Locations}} &\multicolumn{3}{c|}{\textbf{Design Metrics}}  \\\cline{4-6}
		& & & Coverage & Communication quality & Cost  \\\hline
		[20] & \begin{tabular}[c]{p{4.5cm}}Non-dominated sorting genetic algorithm II\centering \end{tabular} & Uniform distribution &\checkmark &\checkmark &     \\\cline{1-6}
		[21] & Genetic algorithm & Not given & &\checkmark & \\\cline{1-6}
		[22] & \begin{tabular}[c]{p{4.5cm}} Progressive construction scheme \centering \end{tabular} & Uniform distribution & &\checkmark & \\\cline{1-6}
		
		[23] & Multilayer tabu search algorithm & Not given &\checkmark &\checkmark & \\\cline{1-6}
		[24] &
		\begin{tabular}[c]{p{4.5cm}} Tabu search Genetic algorithm with opposite-based learning strategy \centering \end{tabular} & Uniform distribution &\checkmark & &\checkmark   \\\cline{1-6}
		[25] &
		\begin{tabular}[c]{p{4.5cm}} Non-negative least squares \centering \end{tabular}& Not given&\checkmark & &  \\\cline{1-6}
		[26] &
		\begin{tabular}[c]{p{4.5cm}} Two-step optimization method \centering \end{tabular}& Not given &\checkmark &\checkmark &  \\\cline{1-6}
		[27] &
		\begin{tabular}[c]{p{4.5cm}}  Multi-objective optimization framework \centering \end{tabular} & Not given &\checkmark &\checkmark &  \\\cline{1-6}
		[28] &
		\begin{tabular}[c]{p{4.5cm}} Two-stage algorithm based on reverse design \centering \end{tabular} & Not given &\checkmark &\checkmark &   \\\cline{1-6}
		This work &
		\begin{tabular}[c]{p{4.5cm}} Improved genetic algorithm \centering \end{tabular} & Poisson point process &\checkmark &\checkmark &  \checkmark

		\\\hline
		
	\end{tabular}
\end{table*}	

   {In fact, the aforementioned basic constellation models have already been applied to LEO satellite IoT \cite{F1}-\cite{F5}.} With the developments of IoT, the optimization of satellite constellations for various applications has received significant attention in recent years. Researchers explored various approaches to enhance the performance, coverage, and reliability of satellite constellations \cite{res2}-\cite{ref25}. For instance, the authors in \cite{res2} proposed a regional satellite constellation design scheme to maximize the target area, and non-dominated sorting genetic algorithm II (NSGA-II) is adopted to address the problem. Moreover, the authors in \cite{res1} proposed a scheme for satellite constellation design by minimizing the system response time, and an optimized algorithm based on genetic algorithm (GA) was proposed for the constellation optimization. Considering the communication delay, the authors in \cite{res3} proposed a progressive construction scheme for satellite constellation design to minimize the end-to-end delay. Considering the user demands, a quality-of-experience (QoE) aware scheme for satellite constellation optimization was proposed in \cite{TS}, and a multilayer tabu search (MLTS) algorithm was designed for obtaining the constellation design scheme to achieve the best QoE. In \cite{ref21}, the authors proposed a tabu search genetic algorithm combined with an opposite-based learning strategy to maximize the network invulnerability benefit. Building upon the need for more unified coverage models, the authors in \cite{ref22} transformed the satellite constellation design into a non-negative least squares problem and derived a unified formula for coverage evaluation. Additionally, literature \cite{ref23} proposed a two-step optimization method to design an asymmetric satellite constellation with resolution requirements for hurricane monitoring in specific areas. Moreover, for emergency observation, the authors in \cite{ref24} investigated the performance metrics of both responsiveness and coverage to approximate the actual emergency conditions, and then proposed a multi-objective optimization framework for satellite constellation design. Furthermore, literature \cite{ref25} proposed a two-stage algorithm based on reverse design for a cross-domain fusion constellation integrating communication, navigation and remote sensing. In this work, we mainly consider the coverage ratio and backhaul capacity for the QoS requirements in LEO satellite constellation design. Specifically, the coverage ratio requirement is used to ensure the coverage of the target area and backhaul capacity requirement is used to guarantee the quality of communication.

   However, there are still several important issues not being addressed in the previous researches. Firstly, the aforementioned works mainly focus on the aspects of the satellite deployment but ignore the terrestrial users and devices or simply assuming that the locations of terrestrial users and devices obey the uniform distribution, which is different from the real situation. {Secondly, most existing studies mainly focused on one or two of the three key design metrics---cost, communication quality, and coverage. Thirdly, the optimization techniques adopted in the existing works have low convergence speed or cannot achieve good performance, which limit the overall performance of the constellation.} Motivated by this, we intend to design a stable and effective satellite constellation optimization scheme with QoS provision. Specifically, the main contributions of this paper are listed as follows:
     \begin{enumerate}
        \item We provide an in-depth performance analysis of LEO satellite IoT constellation. Specifically, we derive out the coverage ratio and transmission rate as a function of constellation parameters and device distribution.
        \item We propose a LEO satellite IoT constellation optimization algorithm by minimizing the total deployment cost while fulfilling QoS requirements in terms of coverage ratio and communication quality.
        \item We analyze the convergence and computational complexity of our proposed algorithm. Moreover, we provide extensive simulation results to verify the effectiveness of the proposed algorithm.
    \end{enumerate}
	For clarity, we summarize the recent works on constellation design, point out their key limitations, and present the solutions of this work in Table I.
	
   The remainder of this paper is organized as follows. In Section II, we introduce the system model of the LEO satellite IoT constellation, including the orbit elements, constellation model, coverage model and channel model. In Section III, we derive out the backhaul capacity expressions and introduce the cost model. Based on these, a cost minimization problem of LEO satellite constellation is formulated and a satellite constellation optimization algorithm is proposed. Next, extensive simulation results are provided in Section IV to verify the effectiveness of the proposed algorithm. Finally, Section V concludes the paper.

   \emph{Notations}: We use bold {lowercase letters and bold uppercase letters} to denote the column vectors and matrices, respectively, $(\cdot)^H$ to denote conjugate transpose, $||\cdot||_F^2$ to denote Frobenius norm of the matrix, $\mathbb{P}\{\cdot\}$ to denote the event probability, $\mathbb{E}\{\cdot\}$ to denote the expectation. $\mathcal{CN}(\mu,\sigma^2)$ denotes the complex Gaussian distribution with mean $\mu$ and variance $\sigma^2$, $\mathbf{I}_M$ denotes the identity matrix of dimension $M$.

\section{System Model}

   We consider the scenario of a LEO satellite IoT constellation as illustrated in Fig. 1, where a large number of LEO
satellites serve a massive number of IoT devices including terrestrial devices and low altitude devices distributed over the world. To satisfy QoS requirements in terms of wireless coverage and transmission rate with a low cost, the satellite constellation architecture is supposed to be designed according to the characteristics of IoT services. In general, the LEO satellites in the constellation are operated in $P$ planes of altitude $h$, denoted by $P=\{1,2,\cdots,P\}$, wherein the orbital plane $p$ consists of $N_p$ LEO satellites, denoted by $\mathcal{N}_P=\{1,2,\cdots,N_P\}$.  In what follows, we first introduce the satellite orbit elements and the constellation model, then present the coverage model of one LEO satellite and the whole satellite constellation, respectively. Finally, we present the channel model in LEO satellite IoT constellation.

    \begin{figure}
    	\centering
    	\includegraphics [width=0.5\textwidth] {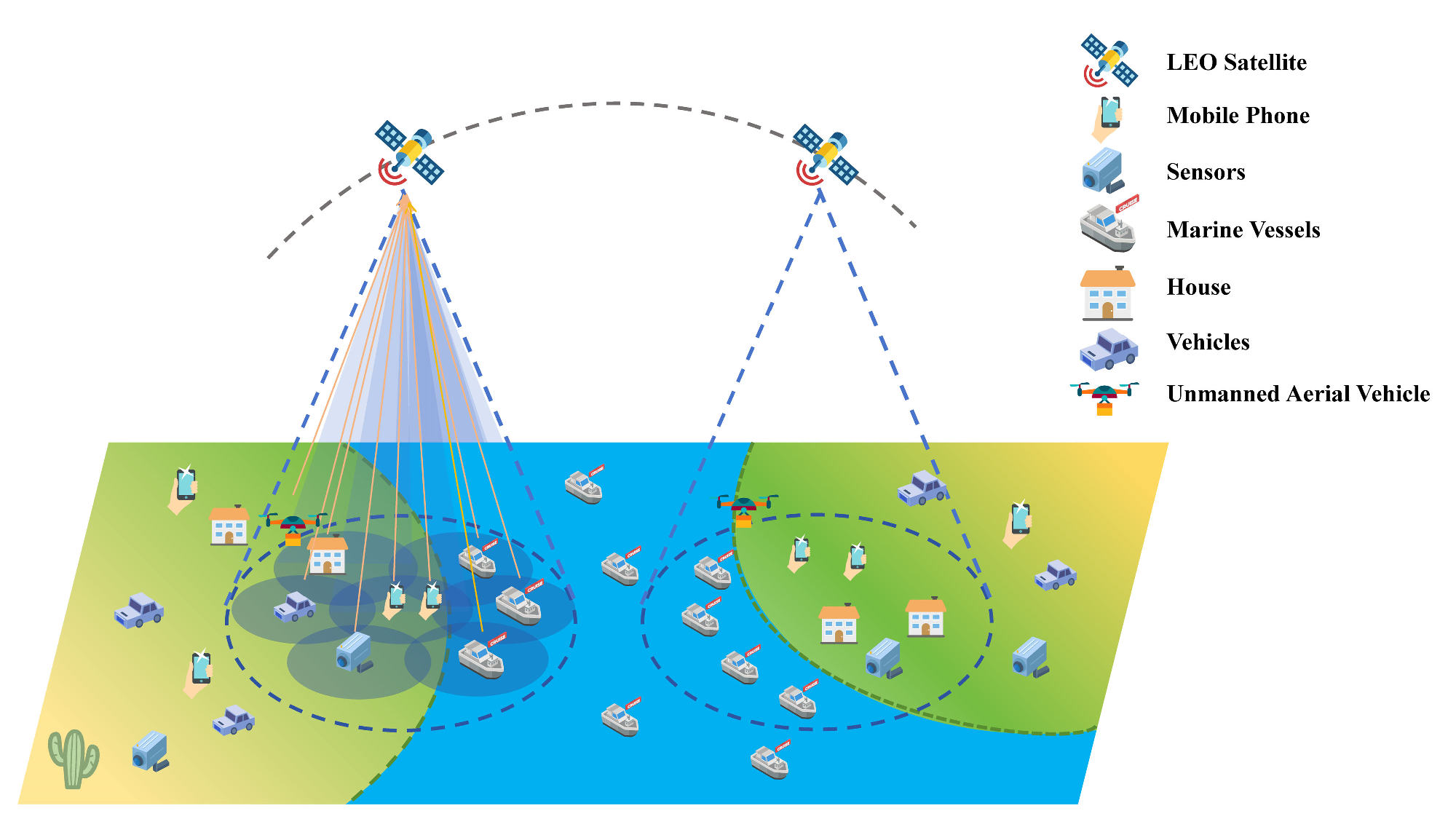}
    	\caption {The communication scenario of LEO satellite IoT, where a large number of LEO satellites
    		serve a massive number of IoT devices distributed over the world.}
    	\label{model}
    \end{figure}

    \begin{figure}
        \centering
        \includegraphics [width=0.45\textwidth] {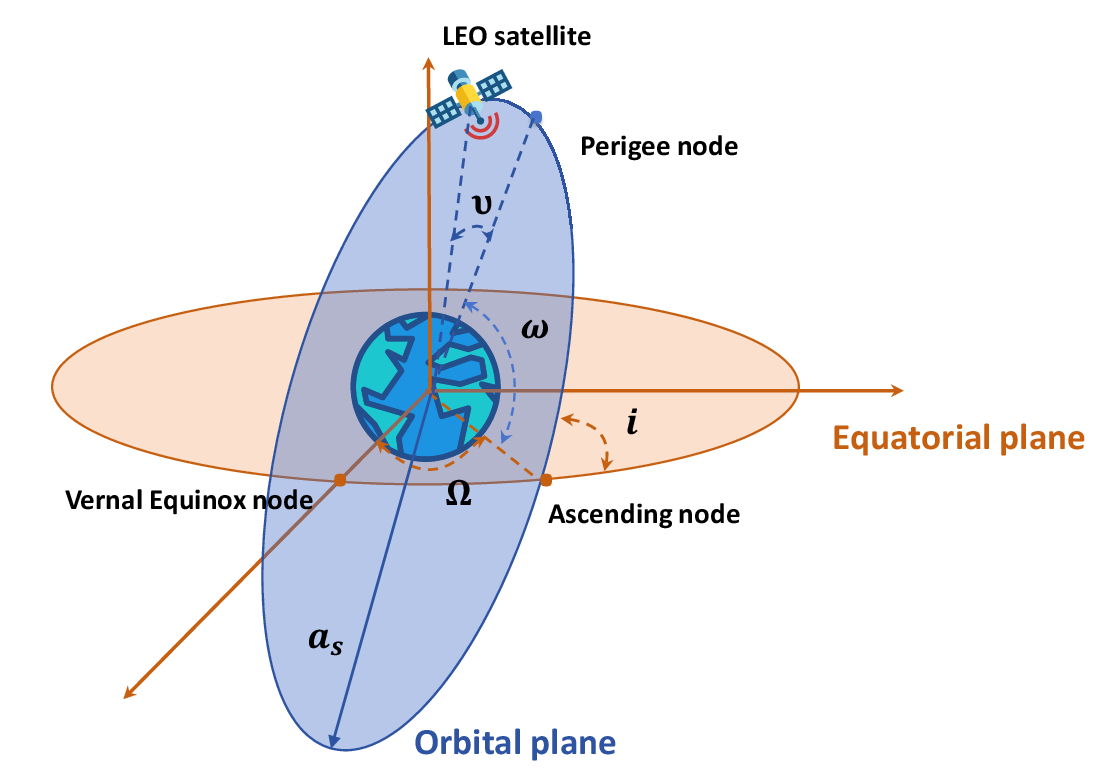}
        \caption {LEO satellite orbit elements.}
        \label{orbit}
    \end{figure}

    \subsection{Orbit Elements and Constellation Model}

    Firstly, we introduce the orbit elements in the LEO satellite IoT constellation. Following previous relevant works \cite{ass1}-\cite{ass3}, we assume that Earth is an ideal sphere and the satellite orbit is circular\footnote{Accounting for Earth’s oblateness and the uneven mass distribution, an oblate spheroid model may be more accurate for Earth shape. For ease of analysis, this paper adopts the commonly used spherical Earth model. In the follow-up researches, a practically inclined elliptical orbit can be adopted for the orbit model.}. As shown in Fig. 2, six orbital elements are defined in the Earth-Centered Inertial (ECI) reference frame to describe the size, shape, and orientation of a LEO satellite’s orbit as follows:
 	\begin{enumerate}
 		
	\item Semi-major $a_s$: Half of the length of the orbit’s major axis. It determines the size of the orbit and directly affects the satellite’s altitude. In circular orbits, it is equal to the radius of the orbit, i.e., $a_s = R_e + h$, where $R_e$ is the radius of Earth and $h$ denotes the satellite altitude.
	\item Eccentricity $e$: The eccentricity of the orbital ellipse.  It determines the shape of the orbit. Given the circular orbit, the eccentricity $e$ is 0.
	\item Inclination angle $i$: Angle measured counterclockwise from the equatorial plane to the satellite orbital plane at the ascending node. It describes the tilt of the orbit relative to Earth’s equatorial plane. {An inclination angle of more than $90^\circ$ indicates a retrograde orbit, where the satellite's motion is opposite to that of Earth’s rotation.}
	\item Right ascension of ascending node $\Omega$: Angle measured counterclockwise from the vernal equinox node to the ascending node in the equatorial plane. It determines the orientation of the orbital plane in space.
	\item Argument of the perigee $\omega$: Angle measured counterclockwise from the ascending node to the perigee, i.e., the point when the satellite operates closest to Earth, measured along the orbital plane. It specifies the location of perigee within the orbit.
	\item True anomaly $\upsilon$: Angle measured counterclockwise from the perigee to the satellite position. It determines the satellite’s position along its orbit at a given time.
	\end{enumerate}
    With these six orbit elements, the position of each satellite can be described accurately. Mathematically, the Cartesian coordinate of each satellite can be expressed as \cite{position} \footnote{This model is valid only under the assumption of a spherical Earth.}
    \begin{equation}\label{cor}
    \left(
    \begin{aligned}
        x_{\mathrm{sat}} & \\
        y_{\mathrm{sat}} & \\
        z_{\mathrm{sat}} & \\
    \end{aligned}\right) = (h+R_e)\left(
    \begin{aligned}
        \mathrm{cos}(\omega+\upsilon)\mathrm{cos}\Omega&-\mathrm{sin}(\omega+\upsilon)\mathrm{sin}\Omega\mathrm{cos}i  \\
        \mathrm{cos}(\omega+\upsilon)\mathrm{sin}\Omega&+\mathrm{sin}(\omega+\upsilon)\mathrm{cos}\Omega\mathrm{cos}i   \\
        \mathrm{sin}(&\omega+\upsilon)\mathrm{sin}i    \\
    \end{aligned}\right)
    \end{equation}
    {For the constellation model, the specific topologies of classical constellations, e.g., the polar orbit constellation, the Walker-Delta constellation, and the Flower constellation, have been extensively studied for their capabilities to address particular coverage requirements while maintaining symmetries and making assumptions that aid analysis, deployment, and operation \cite{J1}. Moreover, they are used as parameter sets in both coverage analysis and network performance analysis \cite{F2}, \cite{J2}. Therefore, we adopt Walker-Delta constellation as the constellation model for optimization in this work.}

     Specifically, for the Walker-Delta constellation, it can be described with three parameters $[N, P, F]$, where $N$, $P$ and $F$ denote the number of satellites per satellite plane, the number of satellite planes and the phase factor, respectively. For the phase factor $F$, it takes a value between 0 to $(P-1)$, which determines the phase difference $\nabla u$ between the corresponding satellites in two adjacent satellite planes, i.e.,
    \begin{equation}\label{dfu}
        \nabla u = \frac{360}{(N* P)}*F.
    \end{equation}

    \subsection{Coverage Model}

    \begin{figure}
        \centering
        \includegraphics [width=0.4\textwidth] {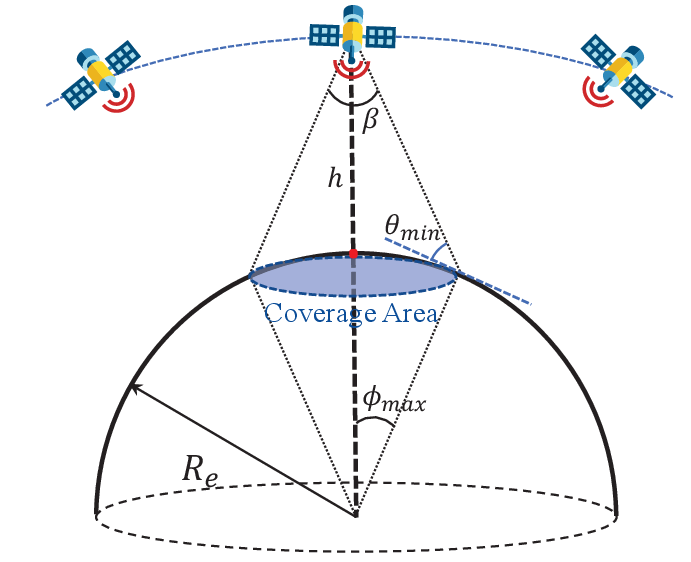}
        \caption {The coverage model of a single satellite.}
        \label{cov}
    \end{figure}

    Wireless coverage is a key factor in the design of LEO satellite IoT constellation. As illustrated in Fig. \ref{cov}, the coverage model of a LEO satellite is determined by the altitude $h$ and the minimum elevation angle $\theta_{\min}$\footnote{ The coverage model used in this work assumes line-of-sight above a certain elevation threshold.}. In this way, the angular radius $\phi_{\max}$, i.e., the angle from the
    point with the minimum elevation angle $\theta_{\min}$ to the projection point of the satellite, can be expressed as
    \begin{equation}\label{phi}
        \phi_{\max} = \arccos\big(\frac{R_e}{R_e+h}\cos\theta_{\min}\big)-\theta_{\min}.
    \end{equation}
    Similarly, the satellite coverage angle $\beta$ can be computed as
    \begin{equation}\label{tensor}
        \beta= 2\cdot \arcsin\big(\frac{R_e}{R_e+h}\cos\theta_{\min}\big).
    \end{equation}
    For a given angular radius $\phi_{\max}$, the coverage area of a LEO satellite can be calculated as
    \begin{equation}\label{nsat}
        S_{\mathrm{sat}} = 2\pi R_e^2(1-\cos\phi_{\max}).
    \end{equation}
    Therefore, the satellite at a higher altitude $h$ can provide wider coverage, while a larger elevation angle $\theta_{\min}$ implies the decrease of coverage.

    In above, we obtained the coverage area of a LEO satellite. Actually, we are more concerned about the coverage ratio of the LEO satellite IoT constellation. To calculate the coverage ratio $\eta_{\mathrm{sat}}$ of the target area served by the LEO satellite IoT constellation, the grid  method is adopted as illustrated in Fig. 4. Specifically, it grids the target area into observation points $\mathcal{G} =\{1, 2, \cdots, G_{\mathrm{map}}\}$. Moreover, the total observation time $T_{\mathrm{total}}$ is divided into multiple time slots. Then, we use the binary variables $o_{\hat{g}}(t)$ to denote whether the observation grid $\hat{g}\in\mathcal{G}$ is covered or not by the satellite IoT constellation at the time slot $t\in\mathcal{T}=\{1, 2,\cdots, T_{\mathrm{total}}\}$, i.e.,
    \begin{equation}\label{gp}
     o_{\hat{g}}(t) = \left\{
    \begin{aligned}
           &1 \quad \mathrm{within}\quad\mathrm{coverage}\\
           &0 \quad \mathrm{beyond}\quad\mathrm{coverage}\\
    \end{aligned}\right..
    \end{equation}
    Based on the location of a satellite under Cartesian coordinates in (\ref{cor}) and the coverage area of a satellite in (\ref{nsat}), we can indicate whether an observation grid is covered as conducted in (\ref{gp}). Then, the coverage ratio of a LEO satellite IoT constellation in time slot $t$ can be calculated as
    \begin{equation}\label{crt}
     \eta_{\mathrm{sat}}(t)=\frac{1}{G_{\mathrm{map}}}\sum_{\hat{g}\in\mathcal{G}}o_{\hat{g}}(t).
    \end{equation}
    Without loss of generality, we use $\eta_{\mathrm{sat}}$ to denote the minimum coverage ratio of the LEO satellite IoT constellation in all time slots, i.e.,
    \begin{equation}\label{crt}
     \eta_{\mathrm{sat}}= \min\{\eta_{\mathrm{sat}}(t)\}, \quad  \forall t.
    \end{equation}
    Based on the above analysis, the calculation of coverage ratio is related to the grid division precision and time division.  Specifically, the coverage ratio is more accurate with higher grid precision and shorter time steps. However, increasing grid precision and shortening time steps inevitably increase computation time \cite{time1}. Thus, it makes sense to select the proper grid precision and time steps. Fortunately, Satellite Tool Kit (STK) is a helpful tool for us to get the precise numerical results including the coverage ratio for satellite constellations.

    \begin{figure}
        \centering
        \includegraphics [width=0.45\textwidth] {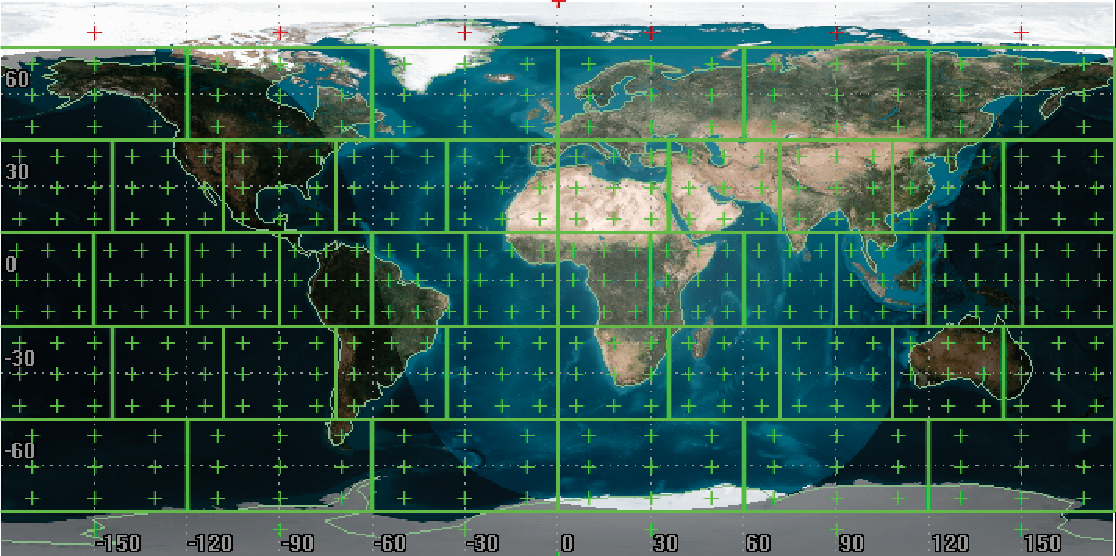}
        \caption {Division of Earth’s surface using the grid method in STK with a grid precision of $10^{\circ}$.}
        \label{grid}
    \end{figure}

    \subsection{Channel Model}
        \begin{figure}
    	\centering
    	\includegraphics [width=0.45\textwidth] {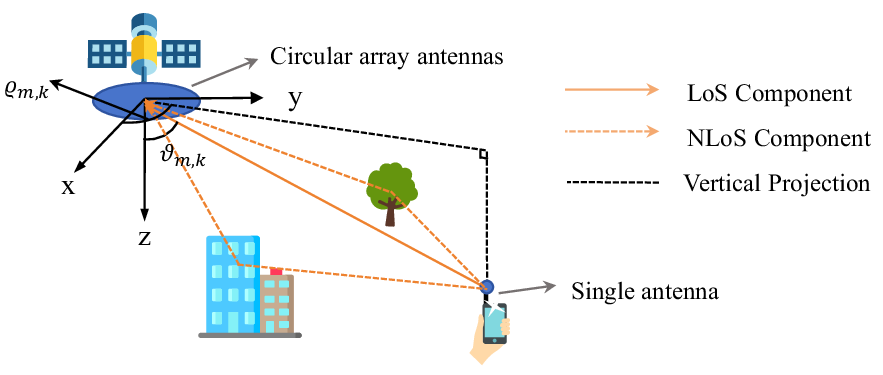}
    	\caption {The channel model between the LEO satellite and IoT devices.}
    	\label{chan}
    \end{figure}
  {According to the signal propagation characteristics of LEO satellite communications \cite{model1}, the LEO satellite channel usually includes two components, namely line of sight (LoS) and non-line of sight (NLoS), as shown in Fig. \ref{chan}.} To be more specific, the channel from the device $k$ to the satellite $m$ can be expressed as\footnote{In this paper, the Doppler frequency shift caused by the motion of LEO satellites is pre-compensated in advance. As the velocity of the satellite can be known from the Ephemeris information, the Doppler frequency shift caused by satellites can be calculated and compensated according to \cite{CFO}.}
    \begin{equation}\label{hmj}
    \begin{aligned}
        \mathbf{h}_{m,k} &= g_{m,k}\bigg(\sqrt{\frac{\lambda_{m,k} }{\lambda_{m,k}+1}}\mathbf{h}^{\text{LoS}}_{m,k}+\sqrt{\frac{1}{\lambda_{m,k}+1}}\mathbf{h}^{\text{NLoS}}_{m,k}\bigg) \\
        &= g_{m,k}\cdot\tilde{\mathbf{h}}_{m,k},\\
    \end{aligned}
    \end{equation}
   where $g_{m,k}$ denotes the channel large-scale fading factor, which can be calculated as
   \begin{equation}\label{gmj}
   	g_{m,k} = \sqrt{ (\frac{c}{4\pi f_cd_{m,k}})^2\cdot G_{\mathrm{dev}}G_\mathrm{sat}\cdot\frac{1}{r_0}},
   \end{equation}
  where $(\frac{c}{4\pi f_cd_{m,k}})^2$ is the free path loss with $c$ being the light speed, $f_c$ being the carrier frequency, and $d_{m,k}$ being the distance between the satellite $m$ and the device $k$ given by
  \begin{equation}\label{dmk}
   	d_{m,k} = -R_e \mathrm{sin} \theta_{m,k}+\sqrt{(R_e \mathrm{sin} \theta_{m,k})^2+h^2+2hR_e},
   \end{equation}
  where $\theta_{m,k}$ denotes the elevation angle of the device $k$ with respect to the satellite $m$. Moreover, $G_{\mathrm{dev}}$ denotes the transmit antenna gain of IoT devices, $G_{\mathrm{sat}}$ denotes the receive antenna gain of satellites, $r_0$ denotes the average rain attenuation coefficient. For the small-scale channel gain $\tilde{\mathbf{h}}_{m,k}$, $\lambda_{m,k}$ denotes the Rician factor of the channel between the satellite $m$ and the device $k$. The LOS component $\mathbf{h}^{\text{LoS}}_{m,k}$ can be modeled as
   \begin{equation}\label{LoS}
   	\begin{aligned}
   	\mathbf{h}^{\text{LoS}}_{m,k} = \sqrt{M_a}\mathbf{a}_{m,k},\\
   \end{aligned}
   \end{equation}
   where $M_a$ denotes the number of satellite antennas, and $\mathbf{a}_{m,k}$ is the array responding vector of antennas. Without loss of generality, each LEO satellite is equipped with a uniform circular array which is widely adopted in current LEO satellite constellations. Thus, the array responding vector $\mathbf{a}_{m,k}$ can be expressed as
   \begin{equation}\label{arv}
   	\begin{aligned}
   		\mathbf{a}_{m,k}(\varphi_{m,k},\varrho_{m,k}) = &\frac{1}{\sqrt{M_a}}
     \big[\exp\{j\varphi_{m,k}\mathrm{cos}(\varrho_{m,k})\},\\
     &\exp\{j\varphi_{m,k}\mathrm{cos}(\varrho_{m,k}-\eta_1)\}, 	\cdots,	\\
     &\exp\{j\varphi_{m,k}\mathrm{cos}(\varrho_{m,k}-\eta_{M_a-1})\}\big],\\
   	\end{aligned}
   \end{equation}
   where $\eta_j = 2\pi j/M_a, j= 0,1,\cdots, M_a$, $\varphi_{m,k}=\frac{\pi d_s f_c}{c}\sin\vartheta_{m,k}$ with $d_s$ being the diameter of circular antenna array equipped on the satellite and $\vartheta_{m,k}$ being the off-axis of the satellite $m$ boresight to the device $k$, and $\varrho_{m,k}$ is the azimuth angle of the device $k$ to the array center of satellite $m$. In addition, $\mathbf{h}^{\text{NLoS}}_{m,k}$ represents the NLoS component of the LEO satellite channel which follows the complex Gaussian distribution, i.e., $\mathbf{h}^{\text{NLoS}}_{m,k}\sim \mathcal{CN}(\mathbf{0}, v_{m,k}^{\mathrm{NLoS}}\mathbf{I}_{M_a})$ with $v_{m,k}^{\mathrm{NLoS}}$ being the variance.

   \section{LEO Satellite Constellation Design}
        In this section, we first introduce the communication model and then develop a cost model for the LEO satellite IoT constellation. Based on these, we formulate a constellation optimization problem by minimizing the cost while providing QoS guarantee in terms of communication quality and coverage ratio, and thereby derive an efficient constellation design algorithm.
   \subsection{Communication Model for LEO Satellite IoT}
   In LEO satellite IoT, due to the bursty characteristics of IoT services, only a small portion of IoT devices are active. In a certain time slot, active IoT devices simultaneously transmit their data sequences of length $L$ over the same time-frequency resource block \cite{D12}, \cite{D22}. Thus, the received signal $\mathbf{Y}_m\in\mathbb{C}^{L\times M_a}$ at the satellite $m$ can be expressed as
   \begin{equation}\label{signal}
   	    \mathbf{Y}_m = \alpha_k\underbrace{\sqrt{\xi_k}\mathbf{s}_k\mathbf{h}_{m,k}^H}_{\text{Desired signal}}+\underbrace{\sum_{\begin{subarray}{c}k'\in\mathcal{D}_{m}\\k'\neq k\end{subarray}}\alpha_{k'}\sqrt{\xi_k}\mathbf{s}_{k'}\mathbf{h}_{m,k'}^H}_{\text{Inter-device Interference}}+\underbrace{\mathbf{n}_m}_{\text{Noise}},
   \end{equation}
   where $\mathcal{D}_{m}$ denotes the interference area of the satellite $m$, $\alpha_k $ is the activation indicator of the $k$-th device with $\alpha_k =1 $ if the device $k$ is active, otherwise, $\alpha_k = 0$. Let $\varepsilon_k$ represent the active probability of the device $k$, then we have
   \begin{equation}\label{pb}
   	\begin{cases}
   		\mathbb{P}\{\alpha_k =1\} = \varepsilon_k\\
   		\mathbb{P}\{\alpha_k =0\} = 1-\varepsilon_k.\\
   	\end{cases}
   \end{equation}
  Moreover, $\xi_k$ denotes the transmit power of the device $k$, $\mathbf{s}_k$ is the transmitted data sequence, which is Gaussian distributed with unit norm, and $\mathbf{n}_m$ denotes the additive white Gaussian noise (AWGN) with variance $\sigma_0^2$. Hence, the signal to interference plus noise ratio (SINR) for the link between satellite $m$ and device $k$ can be computed as
   \begin{equation}\label{SINR}
   	    \Gamma_{m,k} = \frac{||\sqrt{\xi_k}\mathbf{s}_k\mathbf{h}_{m,k}^H||_F^2}{ I_{m,k} +\sigma_0^2},
   \end{equation}
   where $I_{m,k}$ is the interference for the link, which can be calculated as
   \begin{equation}\label{signal}
   	    I_{m,k} = \sum_{\begin{subarray}{c}k'\in\mathcal{D}_{m}\\k'\neq k\end{subarray}}\alpha_{k'}||\sqrt{\xi_{k'}}\mathbf{s}_{k'}\mathbf{h}_{m,k'}^H||_F^2.
   \end{equation}
    In the satellite constellation design, we are more concerned with the overall performance of the system rather than that of a single device or single satellite. Hence, we analyze the overall performance according to the distribution of IoT devices. Without loss of generality, the IoT devices are assumed to be distributed as a homogeneous Poisson point process (HPPP) $\Phi$ with device density $\lambda$ devices/km$^2$ \cite{cova1}. Based on this, we can obtain the following theorem:

    \emph{Theorem 1: }  The average interference of the link between the device $k$ and the satellite $m$ can be calculated as
   \begin{equation}\label{einf}
    	\mathbb{E} [I_{m,k}] = \xi\varepsilon L M_a \lambda (\frac{c}{4\pi f_c})^2\cdot G_{\text{sat}}G_{\text{dev}}\cdot \frac{1}{r_0}	2\pi R_e^2\ln(2\frac{R_e}{h}+1).
   \end{equation}
   \begin{IEEEproof}
       See Appendix A.
   \end{IEEEproof}
   Further, with the SINR in (\ref{SINR}), the transmission rate for the link from the device $k$ to the satellite $m$ is given by
   \begin{equation}\label{SINR2}
   	R_{m,k} = B_w\log_2(1+\Gamma_{m,k}),
   \end{equation}
   where $B_w$ is the channel bandwidth. To calculate the average data rate $\mathbb{E} [R_{m,k}]$, we provide the following theorem:

   \emph{Theorem 2: } The term $\mathbb{E} [\log_2(1+\Gamma_{m,k})]$ can be computed as
   \begin{equation}\label{av_rate1}
   \begin{aligned}
   	\Xi=\mathbb{E} [& \log_2(1+\Gamma_{m,k})]= \frac{\pi (R_e+h)}{S_{\mathcal{A}_k}\ln 2\cdot R_e} \Big[\ln(1+\Psi d_m^{-2})d_m^{2}\\
                &+\Psi \ln(d_m^{2}+\Psi)-\ln(1+\Psi h^{-2})d_m^{2}-\Psi \ln(h^{2}+\Psi) \Big] \\
    \end{aligned},
   \end{equation}
   where $d_m$ denotes the maximum communication distance of
satellite $m$. According to (\ref{dmk}), $d_m$ can be given by
  \begin{equation}\label{dmk2}
   	d_{m} = -R_e \mathrm{sin} \theta_{m}+\sqrt{(R_e \mathrm{sin} \theta_{m})^2+h^2+2hR_e}.
   \end{equation}
      \begin{IEEEproof}
       See Appendix B.
   \end{IEEEproof}

   Thereby, the average transmission rate for the link between device $k$ and satellite $m$ can be given by
   \begin{equation}\label{av_rate2}
   	\mathbb{E} [R_{m,k}]=B_w\cdot\mathbb{E} [\log_2(1+\Gamma_{m,k})]=B_w\Xi.
   \end{equation}
In the LEO satellite IoT constellation, one IoT device may be served by multiple satellites due to multiple-satellite coverage. To calculate the total backhaul capacity of device $k$, we define the set of serving satellites for the device $k$ as $\mathcal{A}_k = \{m|\theta_{m,k}\geq \theta_{\min}\}$, where $\theta_{m,k}$ is the elevation angle from the device $k$ to the satellite $m$. Therefore, the total backhaul capacity of the device $k$ can be calculated as
    \begin{equation}\label{capacity}
   	C_{k} = \sum_{m\in \mathcal{A}_k} R_{m,k}.
   \end{equation}
   It is assumed that the average number of satellites in $\mathcal{A}_k$ is $\bar{N} ^{\mathrm{sat}}_k$, then the average total backhaul capacity $\mathbb{E}[C_k]$ can be calculated as
    \begin{equation}\label{av_ck}
   	\mathbb{E}[C_k] = \sum_{m\in \mathcal{A}_k} \mathbb{E} [R_{m,k}] = \bar{N} ^{\mathrm{sat}}_k\cdot \mathbb{E} [R_{m,k}] .
   \end{equation}

\subsection{Cost Model}
    The construction cost of LEO satellite constellations is crucial for enabling efficient planning, budgeting, and the promotion of sustainable operations. In general, the costs associated with the space segment significantly contribute to the overall expenses of LEO satellite constellations. {Therefore, we first analyze the composition of the costs of the space segment \cite{cost}-\cite{cost4}.}\footnote{We adopt the cost model proposed in \cite{cost} and \cite{cost2} and update the coefficients of the model according to \cite{cost3} and \cite{cost4}. Notice that the proposed cost model is an approximation rather than a generalized rule, the coefficients in the model can be replaced according to the costs of different satellite companies.} {Specifically, the costs of the space segment can be divided into manufacturing cost, launch cost, and insurance cost, respectively.} For the manufacturing cost of a LEO satellite, it can be calculated as
    \begin{equation}\label{manu}
    	\varpi_{\text{manu}} = 0.00185\cdot W_{\mathrm{sat}},
    \end{equation}
    where $W_{\mathrm{sat}}$ denotes the weight of a LEO satellite. For the launch cost, it can be modeled as
   \begin{equation}\label{lanuch}
		\varpi_{\text{launch}} = 0.000166\cdot W_{\mathrm{sat}}\cdot (h/1.609)^{0.43}.
	\end{equation}
    Finally, for the insurance cost, it can be estimated based on the manufacturing cost and launch cost as
    \begin{equation}\label{ins}
    \varpi_{\text{ins}} = \beta_{\mathrm{ins}} \cdot (\varpi_{\text{manu}}+\varpi_{\text{launch}} ),
    \end{equation}
    where $\beta_{\mathrm{ins}}$ denotes the insurance ratio factor. Therefore, the total costs of space segment for a LEO satellite constellation are the summation of these three parts as follows
    \begin{equation}\label{ins}
    \varpi_{\text{space}} = N\cdot P(\varpi_{\text{manu}}+\varpi_{\text{launch}}+\varpi_{\text{ins}}).
    \end{equation}
    It is clear that the costs of the satellite constellation are influenced by the number of LEO satellites and the altitude of the deployed constellation. In this context, we aim to design a LEO satellite IoT constellation by minimizing the total cost under QoS constraints.

   \subsection{Problem Formulation}
        To balance the construction cost and the practical utilization of the desired LEO satellite IoT constellation, we formulate a cost minimization problem while satisfying the QoS requirements under the Walk-Delta architecture. Particularly, the design of LEO satellite IoT constellation can be formulated as the following optimization problem
   \begin{subequations}
   	\begin{eqnarray}
    Q1:~	\underset{N, P,h,i ^{\mathrm{sat}}}{\mathop{\text{minimize}}}\,\!\!&&\!\!\!\varpi_{\text{space}}\label{OP1obj}\\
   		\textrm{s.t.}&&\!\!  \eta_{\mathrm{sat}}\geq\eta_{th}, \label{OP1st1}\\
      &&\!\! C_k^t\geq C_{th}, \quad \forall t, k\label{OP1st2}\\
   	                 &&\!\!	 h_{\mathrm{min}}\leq h\leq h_{\mathrm{max}},	\label{OP1st3}\\
   	                 &&\!\!	 P_{\mathrm{min}}\leq P\leq P_{\mathrm{max}},	\label{OP1st4}\\
   	                 &&\!\!	 N_{\mathrm{min}}\leq N\leq N_{\mathrm{max}},\label{OP1st5}\\
   	                 &&\!\!	 i ^{\mathrm{sat}}_{\mathrm{min}}\leq i ^{\mathrm{sat}}\leq i ^{\mathrm{sat}}_{\mathrm{max}}, \label{OP1st6}
   	\end{eqnarray}
   \end{subequations}
    where $C_k^t= \mathbb{E}[C_k](t)$ denotes the average backhaul capacity $\mathbb{E}[C_k]$ in time slot $t$, $C_{th}$ denotes the backhaul capacity threshold, $\eta_{\mathrm{sat}}$ denotes the coverage ratio, and $\eta_{th}$ denotes the coverage ratio requirement.  Specifically, the constraint (\ref{OP1st1}) is the coverage ratio constraint, the constraint (\ref{OP1st2}) is the communication quality constraint, and the constraints (\ref{OP1st3})-(\ref{OP1st6}) are optimization ranges for the constellation parameters of constellation altitude $h$, number of orbit planes $P$, number of satellites per plane $N$ and inclination angle $i^{\mathrm{sat}}$, respectively. {Moreover, as phase factor $F$ is mainly used to avoid satellite collision, the phase factor is not optimized in the problem \cite{D2}.}

    For such an optimization problem, it is difficult to handle the constraints (\ref{OP1st1}) and (\ref{OP1st2}) directly due to the coverage probability and coverage repetitions. Moreover, these terms are intractable to get a closed-form solution. To this end, according to (\ref{av_ck}) and constraint (\ref{OP1st2}), the minimum serving satellites $\bar{N} ^{\mathrm{sat}}_{th}$ for the device $k$ can be calculated as
    \begin{equation}\label{crt}
     \bar{N} ^{\mathrm{sat}}_{th}= \frac{C_{th}}{B_w\Xi}.
    \end{equation}
    Similarly, the minimum serving satellites $\bar{N} ^{\mathrm{sat}}_{\min}$  can be obtained by measuring the coverage ratio via STK \cite{res2}. {Then the constraint (\ref{OP1st2}) is transformed to satisfy $\bar{N} ^{\mathrm{sat}}_{\min}\geq \bar{N} ^{\mathrm{sat}}_{th}$.} Since the formulated problem is a non-linear, non-convex, and discrete complex combined optimization problem, classical optimization techniques based on continuous convex function cannot be employed directly. Fortunately, intelligent optimization techniques can be applied to handle these problems for its low computational complexity and fast convergence speed. Therefore, we develop an intelligent optimization algorithm based on a GA framework to solve the above optimization.

\subsection{Algorithm Design}
   In this subsection, we design a constellation optimization algorithm based on the GA framework to handle the problem. In general, the GA framework includes the following steps: {1) Chromosomes coding, which constructs a data structure for  the optimization parameters. 2) Population initialization, which initializes the chromosomes for the initial population, a good method of population initialization can accelerate the convergence speed.} 3) Fitness function design, which determines the direction of optimization. 4) Operator design, which generates possible solutions for the best fitness value. 5) Next generation selection, which chooses the best chromosomes to form the next population for the algorithm. In the following, we will introduce these steps in detail.

   \emph{1) Chromosomes Coding:} Upon the GA framework, the first step is to code the chromosomes. According to the characteristics of the formulated optimization problem, the coding vector for a chromosome is set as
    \begin{equation}\label{cod1}
        \mathbf{x} = [h, P, N, i^{\mathrm{sat}}].
     \end{equation}
    In addition, we adopt the real coding for the chromosome coding.

    \emph{2) Population Initialization:} {For the initial population generation, we use the random generation method to generate the initial population. Specifically, the chromosomes for the initial population are randomly generated
between the low bound $\mathbf{l}= [h_{\min}, P_{\min}, N_{\min}, i ^{\mathrm{sat}}_{\min}]$ and the upper bound $\mathbf{u}= [h_{\max}, P_{\max}, N_{\max}, i ^{\mathrm{sat}}_{\max}]$.}

    \emph{3) Fitness Function Design:} To evaluate the quality of potential solutions, it is necessary to design a fitness function based on both objective and constraints. For addressing constraint, the penalty function method is widely applied, with a prevalent model being the amalgamation of the original objective function and the penalty function, which can be expressed as
    \begin{equation}\label{pen1}
        F(x) = f(x)+r\cdot p(x),
     \end{equation}
     where $f(x)$ is the objective function, $p(x)$ is the constraint  function, and $r$ is the penalty factor. It can be easily found in (\ref{pen1}) that choosing a proper penalty factor for the problem is important. Specifically, if the penalty factor is too large, the problem will fail into a local optimal solution. Else if the penalty factor is too small, the constraints may not be satisfied. To overcome this challenge, we propose a novel penalty function design method. Firstly, we introduce the concepts of objective satisfaction function and constraint satisfaction function as below:

     \emph{Definition 1}:  The objective satisfaction function $f_s(x)$ is defined as
     \begin{equation}\label{def1}
     	f_s(x) = \frac{f_{\mathrm{max}} - f(x)}{f_{\mathrm{max}} - f_{\mathrm{min}}},
     \end{equation}
     where $f_{\mathrm{max}}$ and $f_{\mathrm{min}}$ are the maximum value and minimum value of $f(x)$, respectively.

     \emph{Definition 2}:  The constraint satisfaction function $p_s(x)$ is defined as
     \begin{equation}\label{def2}
	     p_s(x) = \frac{p_{\mathrm{max}} - p(x)}{p_{\mathrm{max}}},
     \end{equation}
      where $p_{\mathrm{max}}$ is the maximum value of $p(x)$.

      With these two satisfaction functions, we can evaluate the quality of the solutions. In particular, we define a new penalty function based on the two kinds of satisfaction functions as
     \begin{equation}\label{def3}
      	F_s(x) = f_s(x)\cdot p_s(x)^{\frac{m_\mathrm{fe}}{N_\mathrm{totp}}(\alpha_1 -\alpha_2\frac{1}{n_\mathrm{it}})},
      \end{equation}
      where $\alpha_1$ and $\alpha_2$ are hyperparameters to adjust the level of punishments, $m_\mathrm{fe}$ and $N_\mathrm{totp}$ denote the number of infeasible solutions and total solutions of genetic algorithm, respectively, and $n_\mathrm{it}$ is the current number of evolution. Comparing the proposed model (\ref{def3}) with the classical model (\ref{pen1}), we adjust the penalty factor by utilizing the information of the current population rather than a statistic constant. {Specifically, we introduce the ratio of the number of the infeasible solutions to the total solutions $\frac{m_\mathrm{fe}}{N_\mathrm{totp}}$ in the penalty factor. When there are more infeasible solutions, the model will increase the penalties. If there are more feasible solutions, the model will reduce the penalties and correspondingly increase the value of objective function. Hence, the penalty factor is readily manageable. More importantly, the designed penalty factor promotes the search towards a better solution, leading to a faster convergence speed.} However, if we adopt the classical model in (\ref{pen1}) for penalty, it is difficult to distinguish the feasible solutions and the infeasible ones, but the proposed penalty model can address this challenge by utilizing the product form of two satisfaction functions.

      To apply the proposed penalty function in problem $Q1$, the objective function $f_1(\mathbf{x})$ and constraint functions $p_1(\mathbf{x})$ and $p_2(\mathbf{x})$ should be written as
    \begin{equation}\label{01}
     	f_1(\mathbf{x}) = \varpi_{\text{space}},
     \end{equation}
    \begin{equation}\label{02}
     	p_1(\mathbf{x}) = \Phi(\eta_{\mathrm{sat}}-\eta_{th}),
     \end{equation}
    \begin{equation}\label{03}
     	p_2(\mathbf{x}) = \Phi(\bar{N} ^{\mathrm{sat}}_{\min}-\frac{C_{th}}{B_w\Xi}),
     \end{equation}
    where $\Phi(x) = \max\{x,0\} $ is used to measure the degree of violation of the constraints. To calculate the corresponding satisfaction functions, we introduce $f_1^{\min}$, $f_1^{\max}$,  $p_1^{\max}$, and $p_2^{\max}$ to denote the minimum and maximum values of objective function and constraint functions (\ref{01}) - (\ref{03}) in the population, respectively. As a result, the satisfaction functions of (\ref{01})-(\ref{03}) can be constructed as
    \begin{equation}\label{s01}
     	f_{s1}(\mathbf{x}) = \frac{f_1^{\max}-f_1(\mathbf{x})}{f_1^{\max}-f_1^{\min}},
     \end{equation}
    \begin{equation}\label{s02}
     	p_{s1}(\mathbf{x}) = \frac{p_1^{\max}-p_1(\mathbf{x})}{p_1^{\max}},
     \end{equation}
    \begin{equation}\label{s03}
     	p_{s2}(\mathbf{x}) = \frac{p_2^{\max}-p_2(\mathbf{x})}{p_2^{\max}}.
     \end{equation}
    In this way, the proposed penalty function for problem $Q1$ can be expressed as
    \begin{equation}\label{fit}
      	F_s(\mathbf{x}) = f_{s1}(\mathbf{x})\cdot \big(p_{s1}(\mathbf{x}) p_{s2}(\mathbf{x})\big)^{\frac{m_\mathrm{fe}}{N_\mathrm{totp}}(\alpha_1 -\alpha_2\frac{1}{n_\mathrm{it}})},
    \end{equation}
    which serves as the fitness function to evaluate the quality of potential solution.

    \emph{4) Operation Design:} The fourth step of the GA framework is to design the GA operators, including crossover operator and mutation operator. To improve the ability of the algorithm to find optimal solutions, we add the information of the optimal individual in the crossover operator. Specifically, we use $\mathbf{x}_\mathrm{best}$ to denote the best individual in the $n_\mathrm{it}$-th population $\mathcal{P}(n_\mathrm{it})$. Then, we choose $N_1$ individuals according to roulette wheel selection with their fitness to form a parent set $\mathcal{P}_{pa}$. For the parents $\mathbf{x}_{p,1}$ and $\mathbf{x}_{p,2}$, two new individuals are breed by the following crossover operator
    \begin{equation}\label{cross1}
    \begin{aligned}
        \mathbf{x}_{c,1} = &\frac{r_{11}}{r_{11}+r_{12}+r_{13}}\mathbf{x}_{p,1}+\frac{r_{12}}{r_{11}+r_{12}+r_{13}}\mathbf{x}_{p,2}\\
        &+\frac{r_{13}}{r_{11}+r_{12}+r_{13}}\mathbf{x}_\mathrm{best},\\
    \end{aligned}
    \end{equation}
    \begin{equation}\label{cross2}
    \begin{aligned}
        \mathbf{x}_{c,2} =& \frac{r_{21}}{r_{21}+r_{22}+r_{23}}\mathbf{x}_{p,1}+\frac{r_{22}}{r_{21}+r_{22}+r_{23}}\mathbf{x}_{p,2}\\
        &+\frac{r_{23}}{r_{21}+r_{22}+r_{23}}\mathbf{x}_\mathrm{best},\\
    \end{aligned}
    \end{equation}
     where $r_{ij}(i=1,2;j=1,2,3)$ are random values in the range 0 to 1, $\mathbf{x}_{c,1}$ and $\mathbf{x}_{c,2}\in \mathcal{P}_1(n_\mathrm{it}) $ are the child solutions and $\mathcal{P}_1(n_\mathrm{it})$ is the child generation set.

     For the mutation operator, the algorithm is expected to have a relatively large search range at the beginning of the evolution to preserve the diversity of the population, while at the later stages of the evolution it should be guaranteed that there are enough feasible solutions for the algorithm to converge. Motivated by this, we design the mutation operator with a mutation indicator $\theta_\mathrm{mut}\in[0, 1]$ as follows: if $\theta_\mathrm{mut}\leq (\frac{1}{n_\mathrm{it}+1}+1)\varsigma$, the chromosome after mutation $\mathbf{x}_{mu}$ is determined by
    \begin{equation}\label{xmu1}
     	\mathbf{x}_{mu} = \mathbf{x}+\Delta \mathbf{m},
     \end{equation}
    where $\varsigma$ is the threshold of mutation operator, and $\Delta \mathbf{m}= (\Delta m_1, \Delta m_2, \cdots, \Delta m_n)^T$ with $\Delta m_i$ following the Gaussian distribution with zero mean and $\sigma_i^2$, i.e., $\Delta m_i\sim\mathcal{N}(0, \sigma_i^2)$. Otherwise, if $\theta_\mathrm{mut}> (\frac{1}{n_\mathrm{it}+1}+1)\varsigma$, the chromosome after mutation $\mathbf{x}_{mu}$ is set as
        \begin{equation}\label{xmu2}
     (x_{mu})_i = \left\{
    \begin{aligned}
           &x_i +\frac{r_{3, i}}{n_\mathrm{it}}\cdot(u_i-l_i)\quad  r_{4, i}<0.5\\
           &x_i -\frac{r_{3, i}}{n_\mathrm{it}}\cdot(u_i-l_i)\quad  r_{4, i}\geq0.5\\
    \end{aligned}\right. (i=1, 2, 3, 4),
    \end{equation}
    where $r_{3, i}$ and $r_{4, i}$ are random numbers range in $[0, 1]$, $(x_{mu})_i$ and $x_i$ respectively denotes the $i$-th element of $\mathbf{x}_{mu}$ and $\mathbf{x}$, $u_i$ denotes the $i$-th element of upper bound of search space $\mathbf{u}= [h_{\max}, P_{\max}, N_{\max}, i ^{\mathrm{sat}}_{\max}]$, and $l_i$ denotes the $i$-th element of lower bound of search space $\mathbf{l}= [h_{\min}, P_{\min}, N_{\min}, i ^{\mathrm{sat}}_{\min}]$. Consequently, at the beginning of the evolution, the population will be more likely to mutate according to (\ref{xmu1}). While at the end of the evolution, the population will be more likely to mutate according to (\ref{xmu2}) as to perform local search around the mutated chromosome. For convenience, we use $\mathcal{P}_2(n_\mathrm{it})$ to denote the population after mutation.

\begin{figure*}
        \centering
        \includegraphics [width=0.6\textwidth] {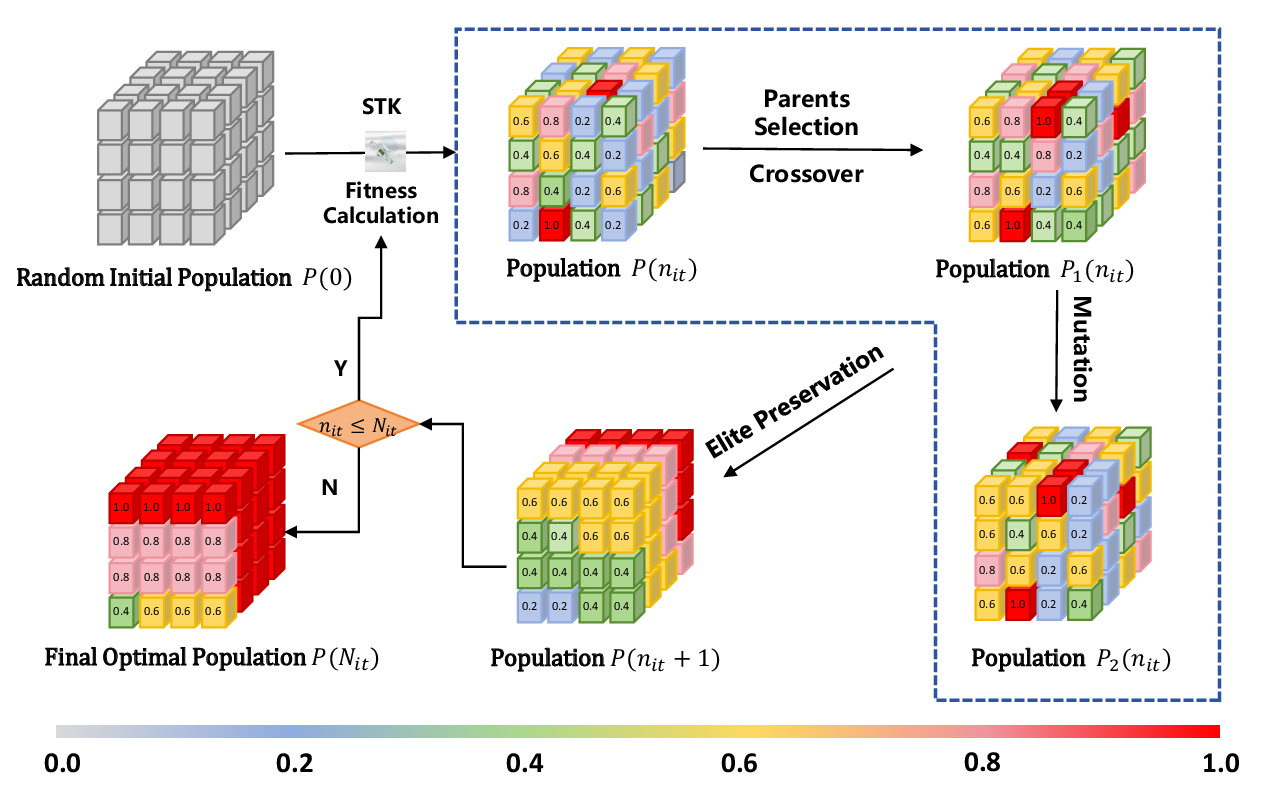}
        \caption { Simplified flow illustration of the proposed algorithm.}
        \label{alg}
    \end{figure*}

    \emph{5) Next Generation Selection:} Finally, we adopt the elite preservation method to form the population of next generation. In other words, we choose the best $N_\mathrm{totp}$ individuals in the set $\mathcal{P}(n_\mathrm{it})\cup \mathcal{P}_1(n_\mathrm{it})\cup \mathcal{P}_2(n_\mathrm{it})$ to form the $\mathcal{P}(n_\mathrm{it}+1)$. The whole algorithm is summarized in \textbf{Algorithm 1}. {For clarity, we describe the algorithm flow in Fig. \ref{alg}.}

\begin{breakablealgorithm}
	\caption{: Design of LEO Satellite IoT Constellation}
	\label{alg1}
	% \hspace*{0.02in}%{\bf Input:}
	%	$\mathbf{Y},\{\mathbf{A_i}\}_{i=1}^d $ and total iterations T \\
	%  \hspace*{0.02in}{\bf Output:}
	%	$\mathbf{M}_X$
	\begin{algorithmic}[1]
		\STATE{\textbf{Input:}  $f_c$, $G_{\mathrm{dev}}$, $G_{\mathrm{sat}}$, $B_w$, $T$, $r_0$, $\beta$, $M$, $\varepsilon_k$, $L$, $\lambda$, $W _{\mathrm{sat}}$, $\eta_{th}$, $C_{th}$, $\mathbf{u}$, $\mathbf{l}$, $N_\mathrm{totp}$, $n_\mathrm{it}$, $\xi_k$  }
		
		\STATE{\textbf{Initialization}: generate $\mathcal{P}(1)=[\mathbf{x}_1,~\mathbf{x}_2,~\cdots,~\mathbf{x}_{N_\mathrm{totp}}]$ randomly with upper bound $\mathbf{u}$ and lower bound $\mathbf{l}$, iteration index $n_\mathrm{it}=1$;}
		\STATE{
			\textbf{While} $n_\mathrm{it} \leq N_\mathrm{it}$
		
		}
		\STATE{\textbf{Calculate the fitness value of individuals} in  $\mathcal{P}(n_\mathrm{it})$
			according to (\ref{01}) - (\ref{fit});
		}
		\STATE{\textbf{Parents Selection}: Choose the best $N_\mathrm{totp}$ individuals in $\mathcal{P}(n_\mathrm{it})$ by roulette wheel selection with the fitness function calculated by step 4 to form the parents' set $\mathcal{P}_{pa}(n_\mathrm{it})$.
		}
		\STATE{\textbf{Crossover:} Randomly choose two individuals in $\mathcal{P}_{pa}(n_\mathrm{it})$ and the best individual to perform the crossover operators according to (\ref{cross1}) and (\ref{cross2}) until getting $N_\mathrm{totp}$ children to form the population $\mathcal{P}_1(n_\mathrm{it})$;}
           ~\STATE{\textbf{Mutation:} For each individual individual in $\mathcal{P}_{1}$, perform the mutation operator as:

           \textbf{For} $s=1:N_\mathrm{totp}$:

            ~~~~~Randomly choose $\theta_\mathrm{mut}\in [0,~1]$;

            ~~~~~\textbf{If} $\theta_\mathrm{mut}\leq (\frac{1}{n_\mathrm{it}+1}+1)\varsigma$

            ~~~~~~~~perform the mutation operator for $\mathbf{x}(s)$ according to (\ref{xmu1});

            ~~~~~\textbf{Else if} $\theta_\mathrm{mut}> (\frac{1}{n_\mathrm{it}+1}+1)\varsigma$

            ~~~~~~~~perform the mutation operator for $\mathbf{x}(s)$ according to (\ref{xmu2});

            The individuals after mutation form the population $P_2(n_\mathrm{it})$;
            }
           \STATE{\textbf{Next Generation Selection:} Calculate the fitness function in the population $\mathcal{P}(n_\mathrm{it})\cup \mathcal{P}_1(n_\mathrm{it})\cup \mathcal{P}_2(n_\mathrm{it})$, and select the best $N_1$ individuals to form the population $\mathcal{P}(n_\mathrm{it}+1)$;}
		\STATE{$n_\mathrm{it}=n_\mathrm{it}+1$;}
            \STATE{\textbf{End while;}}
		\STATE{\textbf{Output:}  the minimum space costs $\varpi_{\mathrm{space}}^{\mathrm{opt}}$, the optimized number of satellites per plane $N_{\mathrm{opt}}$, the optimized number of satellite planes $P_{\mathrm{opt}}$, the optimized altitude $h_{\mathrm{opt}}$, and the optimized inclination angle $i^{\mathrm{sat}}_{\mathrm{opt}}$.

  }
	\end{algorithmic}
\end{breakablealgorithm}

  \subsection{Algorithm Analysis}
      To gain further insights of the proposed algorithm, we analyze its convergence properties and computational complexity.

      \emph{Convergence Analysis:} The proposed algorithm has fast convergence speed due to the special design even with very large constellation parameters. The proof of convergence please refers to Appendix C.

      \emph{Complexity Analysis:} {The proposed algorithm is a kind of heuristic algorithm, which trades off the computational precision for reduced complexity. Moreover, for the candidate solutions generated by the algorithm, parallel computing techniques are adopted to reduce the computational complexity.} For each iteration of our proposed algorithm, the main computational complexity is caused by the calculation of coverage ratio by STK. Therefore, the complexity of our proposed algorithm is $O(N_\mathrm{it}N_\mathrm{totp})$. To verify the effectiveness of the proposed algorithm, we compare it with five typical algorithms which are widely adopted in satellite constellation optimization in Table \ref{compl}. The comparison results show that our proposed algorithm does not increase the computational complexity.

    \begin{table}
	\small
	\centering
	\caption{Computational Complexity Comparison}\label{compl}
	\begin{tabular}{|c|c|}
		\hline
		Proposed algorithm & $N_\mathrm{it} N_\mathrm{totp}$ \\\hline
		Tabu Search  \cite{TS}  & $ N_\mathrm{it}(N_\mathrm{totp}^2+N_\mathrm{totp}+l_{ts})$  \\\hline
        Sine Cosine algorithm  \cite{SCA} & $N_\mathrm{it}N_\mathrm{totp}$ \\\hline
        Particle Swarm optimization  \cite{PSO}  & $N_\mathrm{it}N_\mathrm{totp}$\\\hline
		Grey Wolf Optimization \cite{GWO}   & $N_\mathrm{it}N_\mathrm{totp}$  \\\hline
		Classical GA \cite{GA}   & $N_\mathrm{it} N_\mathrm{totp}$  \\\hline
	\end{tabular}
    \end{table}	

\section{Simulation Results}

    In this section, we provide extensive simulation results to testify the performance of the proposed algorithm for LEO satellite IoT constellation design. According to TR 38.821 \cite{tr}, the main simulation parameters of the satellite constellation are summarized in Table \ref{sys_pa}. Moreover, the range of optimization parameters are set in Table \ref{op_pa}.

\begin{table}
	\small
	\centering
	\caption{Main System Parameters For LEO Satellite IoT Constellation}\label{sys_pa}
	\begin{tabular}{|c|c|}
		\hline
		Parameter & Value\\\hline\hline
		Satellite orbit & LEO \\\hline
		Frequency band & L/S/C  \\\hline
             Earth's radius & 6378.14 km \\\hline
             Target area & [$60^{\circ}$S, $60^{\circ}$N] \\\hline
		Carrier frequency $f_c$& 5 GHz  \\\hline
		Carrier bandwidth $B_w$  & 250 MHz  \\\hline
		Satellite antenna gain $G_{\mathrm{sat}}$  & 17 dBi  \\\hline
		{ IoT device antenna gain $G_{\mathrm{dev}}$}  & { 3 dBi}  \\\hline
		Average rain fading  $r_0$& -2.6 dB \\\hline
		Rician factor $\lambda_{m,k}$ & 10 dB \\\hline
		{ Noise variance  $\sigma_0^2$} & { $-106$ dBm}\\ \hline
		Number of satellite antennas $M_a$ & $16$\\ \hline
            Activity probability $\varepsilon$ & $0.005$\\ \hline
            Density of IoT devices $\lambda$ & $8\times 10^{-5}$\\ \hline
            Average transmit power $\xi$ & $3$ dBw\\ \hline
		Sequence length $L$ & $100$  \\ \hline
		Satellite coverage angle  $\beta$ & $45^\circ$ \\ \hline
		Coverage ratio requirement $\eta_{th}$ & 90\% \\ \hline
		Backhaul capacity requirement $C_{th}$ &$80$ Mbps \\ \hline
            Satellite weight $W _{\mathrm{sat}}$ &$227$ kg \\ \hline
           { Phase Factor $F$} & {1}\\ \hline
            Start time  & 1 Jan 2025 00:00:00.0 \\ \hline
		End time & 1 Jan 2025 23:59:00.0 \\ \hline
		Time step &  60 s \\ \hline
	\end{tabular}
\end{table}	

\begin{table}
	\small
	\centering
	\caption{Optimized Parameters Range of LEO Satellite IoT Constellation }\label{op_pa}
	\begin{tabular}{|c|c|}
		\hline
		Parameter & Value\\\hline\hline
		Minimum altitude $h_{\min}$ & 500 km \\\hline
		Maximum altitude $h_{\max}$ & 1800 km \\\hline
        Minimum number of planes $P_{\min} $ & 4 \\\hline
        Maximum number of planes $P_{\max}$ & 20 \\\hline
        Minimum number of satellites per plane $N_{\min}$ & 4 \\\hline
        Maximum number of satellites per plane $N_{\max}$ & 20 \\\hline
		Minimum inclination angle $i^{\mathrm{sat}}_{\min}$  & $20^\circ$  \\\hline
		Maximum inclination angle $i^{\mathrm{sat}}_{\max}$ & $60^\circ$  \\\hline
	\end{tabular}
\end{table}

    \begin{figure}
        \centering
        \includegraphics [width=0.48\textwidth] {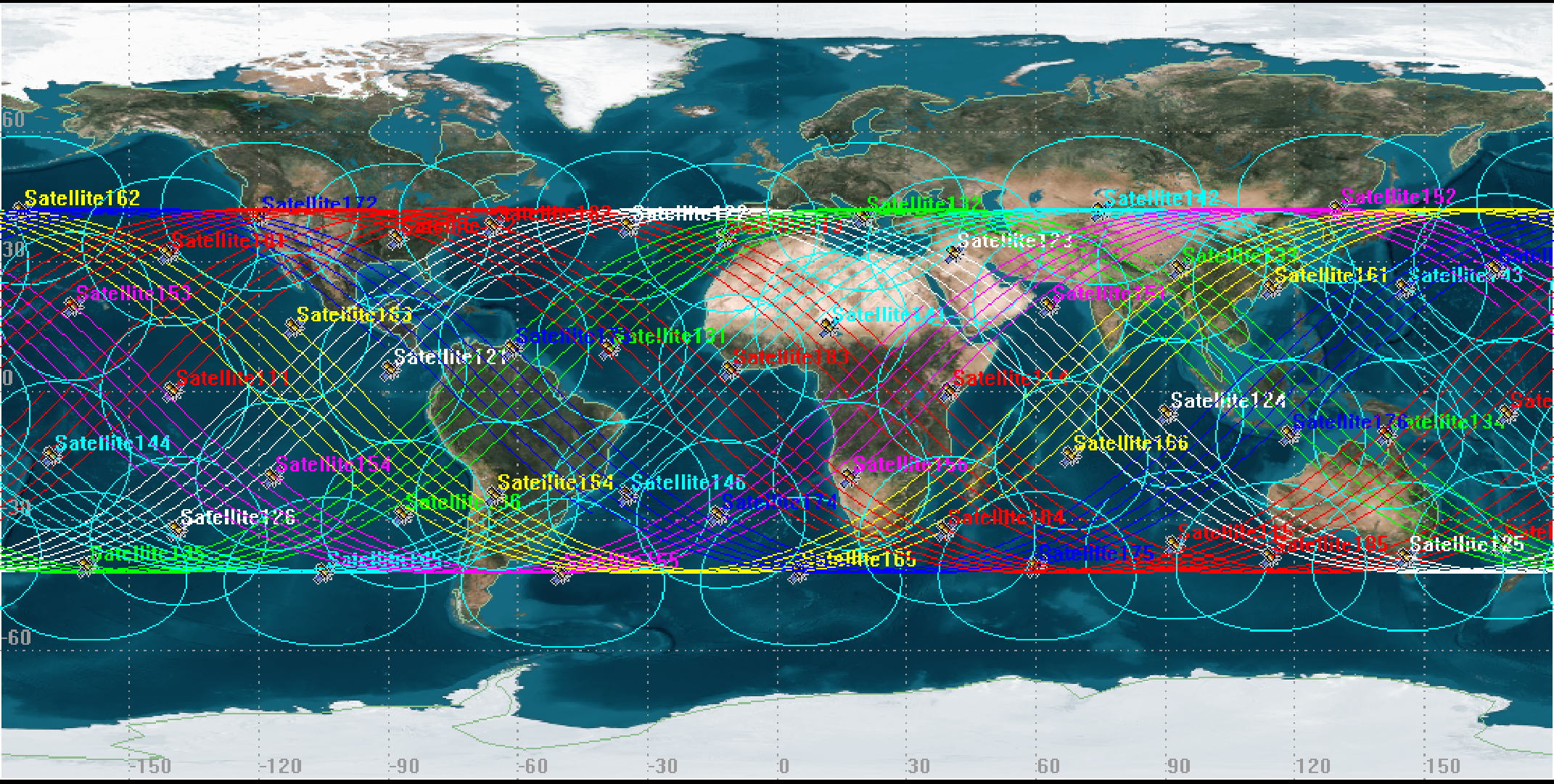}
        \caption { 2D model of the designed LEO satellite IoT constellation.}
        \label{2d}
    \end{figure}

    \begin{figure}
        \centering
        \includegraphics [width=0.45\textwidth] {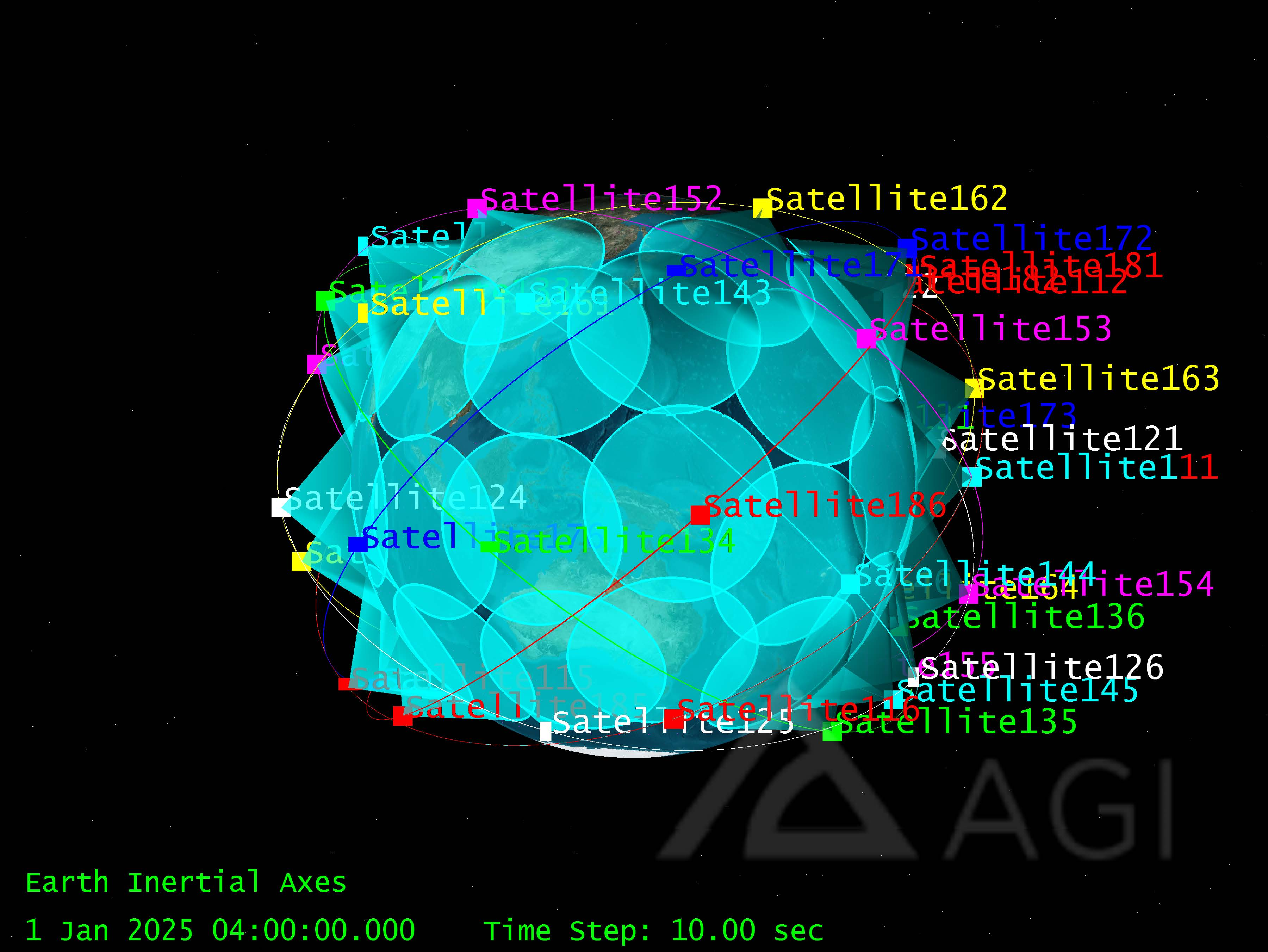}
        \caption {3D model of the designed LEO satellite IoT constellation.}
        \label{3d}
    \end{figure}

    {To visualize the coverage of constellations, Fig. \ref{2d} has provided the two-dimensional (2D) simulation result of the proposed LEO satellite IoT constellation design algorithm under the default simulation parameters of Table \ref{sys_pa} and \ref{op_pa}. Specifically, the optimized number of planes is $P_{\mathrm{opt}} = 6$, the optimized number of satellites per plane is $N_{\mathrm{opt}} = 8$, the optimized altitude is $h_{\mathrm{opt}} = 1589$ km, and the optimized inclination angle $i^{\mathrm{sat}}_{\mathrm{opt}} = 41^\circ$. It can be seen from Fig. \ref{2d} that the proposed LEO satellite IoT constellation design algorithm has provided sufficient coverage for the target area. Moreover, it is presented in Fig. \ref{3d} that the three-dimensional (3D) model of the wireless coverage of the designed LEO satellite IoT constellation, which also confirms the strong coverage performance of the proposed algorithm.}

    \begin{table}
    \scriptsize
	\centering
	\caption{Simulation Parameters for the Proposed Algorithm with Classical GA}\label{Simulation}
	\begin{tabular}{|c|c|c|}
		\hline
		Parameters & Proposed Algorithm & Classical GA \\\hline
		Population size $N_\mathrm{totp}$ & 30 & 30 \\\hline
		  Number of iterations $N_\mathrm{it}$ \centering   & 50 & 50 \\ \hline
		Mutation operator threshold & 0.3 & 0.3 \\ \hline
		Penalty factor $\rho_1$ and $\rho_2$ &  $\backslash$ & 1000, 1000 \\ \hline
		Hyperparameters $\alpha_1$ and $\alpha_2$ & 2, 1 &  $\backslash$\\\hline
	\end{tabular}
\end{table}

    \begin{figure}
        \centering
        \includegraphics [width=0.45\textwidth] {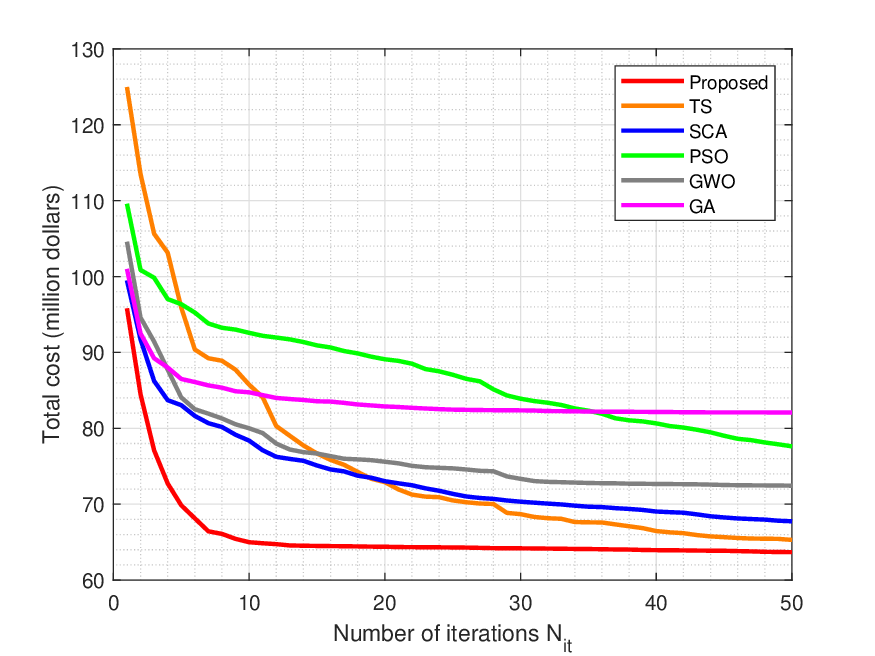}
        \caption {The total cost of LEO satellite IoT constellation with different optimization algorithms in iterations.}
        \label{costit}
    \end{figure}
    Furthermore, to verify the convergence and effectiveness of the proposed algorithm, we compare it with five baseline intelligent algorithms, which are widely adopted in satellite constellation optimization, including classical GA \cite{GA},  Sine Cosine algorithm (SCA) \cite{SCA}, Grey Wolf Optimization (GWO) \cite{GWO}, Particle Swarm optimization (PSO) \cite{PSO}, and Tabu Search (TS) algorithm \cite{TS}. {For the classical GA in \cite{GA}, the crossover operation adopted in classical GA does not take into account information about the current optimal solution, and the mutation operator remains unchanged during the iterations. The simulation parameters of the proposed algorithm and classical GA are listed in Table V.} Moreover, SCA and GWO are two novel swarm intelligent optimization algorithms. For PSO, it is a widely adopted optimization algorithm in which each particle retains memory of its previous best position, making it applicable for many optimization problems. Finally, TS is a metaheuristic local search algorithm by checking its immediate neighbors of the current solution in search of finding an improved solution. {To avoid contingency, statistical analysis is performed for each algorithm. Specifically, the average total cost of each algorithm is evaluated for 200 trials.} As the penalty function in (\ref{fit}) contains several hyperparameters of GA algorithm, it is not suitable for the baseline algorithms to utilize (\ref{fit}) as the fitness function. Therefore, a classical penalty function is adopted for the baseline algorithms as the fitness function, which is expressed as
    \begin{equation}\label{fit2}
      	F_s(\mathbf{x}) = f_{1}(\mathbf{x}) + \rho_1 p_1(\mathbf{x})+\rho_2 p_2(\mathbf{x}),
    \end{equation}
    where $\rho_1$ and $\rho_2$ are two fixed penal factors. In Fig. \ref{costit}, we show the convergence performance of the proposed algorithm alongside five baseline intelligent algorithms. {Evidently, the proposed algorithm demonstrates the fastest convergence among all the algorithms.} Noticeably, due to the special design of the fitness function and algorithm operators, the proposed algorithm has the lowest cost. Therefore, the proposed algorithm is effective in LEO satellite IoT constellation optimization.
    \begin{figure}
        \centering
        \includegraphics [width=0.45\textwidth] {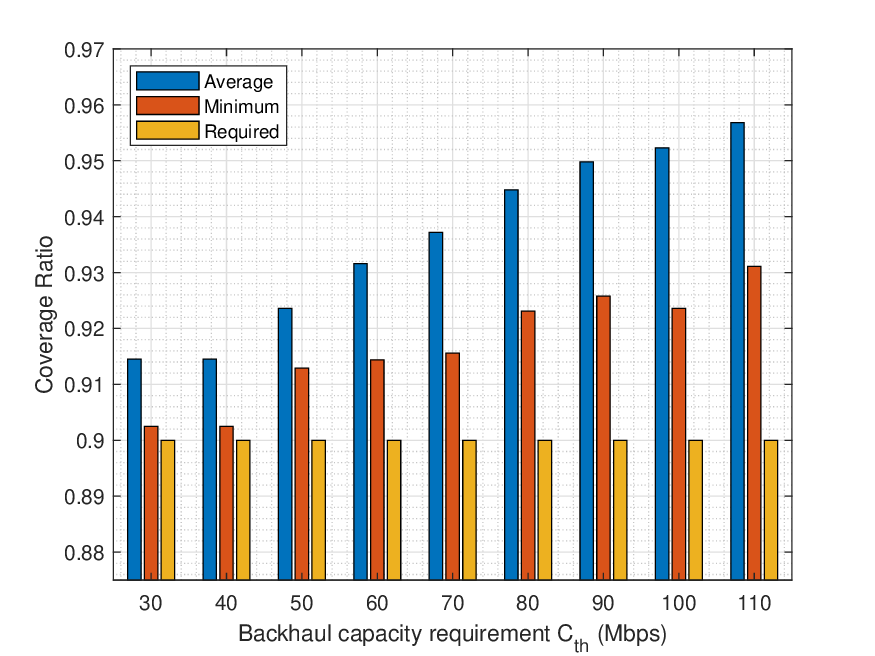}
        \caption { The coverage ratio performance of the proposed LEO satellite IoT constellation under different backhaul capacity requirements.}
        \label{conv1}
    \end{figure}

    \begin{figure}
        \centering
        \includegraphics [width=0.45\textwidth] {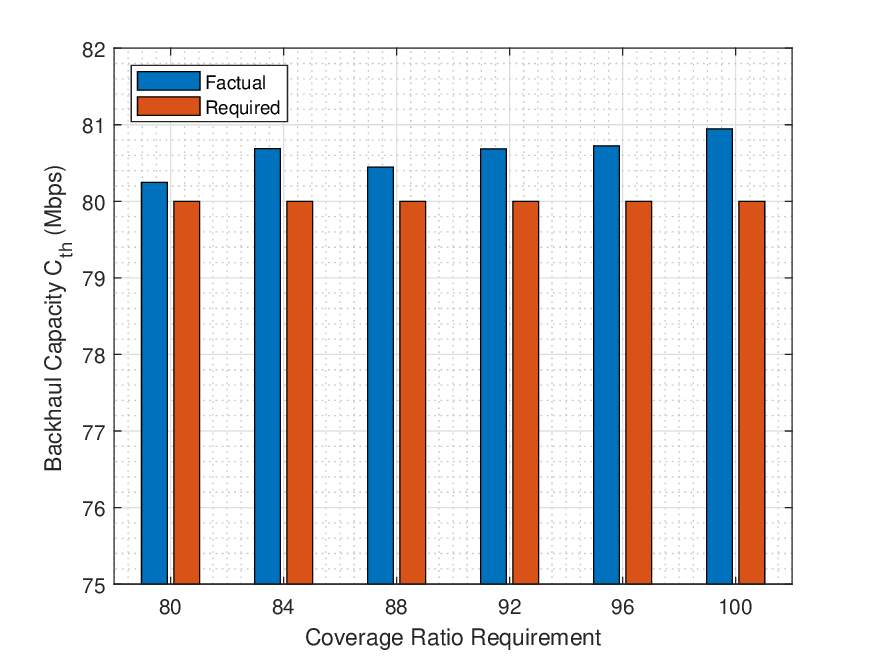}
        \caption { The backhaul capacity performance of the proposed LEO satellite IoT constellation under different coverage ratio requirements.}
        \label{conv1_2}
    \end{figure}

    Moreover, it is essential to verify whether the proposed  algorithm satisfies the both QoS requirements. Hence, we investigate the coverage ratio performance of the proposed algorithm under different backhaul capacity requirements. As illustrated in Fig. \ref{conv1}, the minimum coverage ratio of the proposed algorithm exceeds the desired coverage ratio for a given backhaul capacity. Besides, the average coverage ratio over all time slots grows as the backhaul capacity requirements $C_{th}$ increases. This is because a higher backhaul capacity requirement for IoT devices always needs more LEO satellites to provide coverage for a specific area. {Particularly, the simulation results at 30 Mbps and 40 Mbps demonstrate identical performance, indicating that the proposed constellation is the most cost-effective while meeting the coverage requirements. Additionally, it is illustrated in Fig. \ref{conv1_2} that the proposed algorithm can satisfy the backhaul capacity requirements under different coverage ratio requirements.} Therefore, the proposed algorithm can provide high QoS for the target area.

    \begin{figure}
        \centering
        \includegraphics [width=0.45\textwidth] {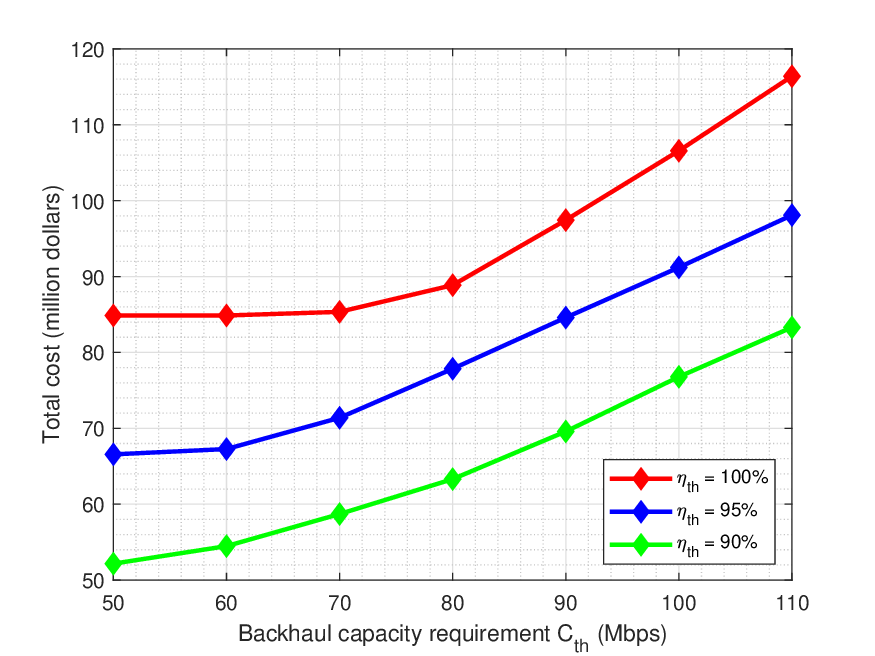}
        \caption { The total cost of the proposed LEO satellite IoT constellation with different backhaul capacity requirements.}
        \label{cov2}
    \end{figure}

   Subsequently, we explore the relationship between the total cost and the QoS requirements. As depicted in Fig. \ref{cov2}, a higher backhaul capacity requirement or coverage ratio requirement will increase the total cost of the LEO satellite IoT constellation. Furthermore, it is also found that the total cost keeps nearly unchanged when $C_{th}$ increases from $50$ Mbps to $60$ Mbps at the coverage ratio requirement of $100\%$, {since the constellation with seamless coverage has achieved the backhaul capacity more than 60 Mbps. }In this way, we should balance the total cost and the QoS requirements when designing the LEO satellite IoT constellation.

    \begin{figure}
        \centering
        \includegraphics [width=0.45\textwidth] {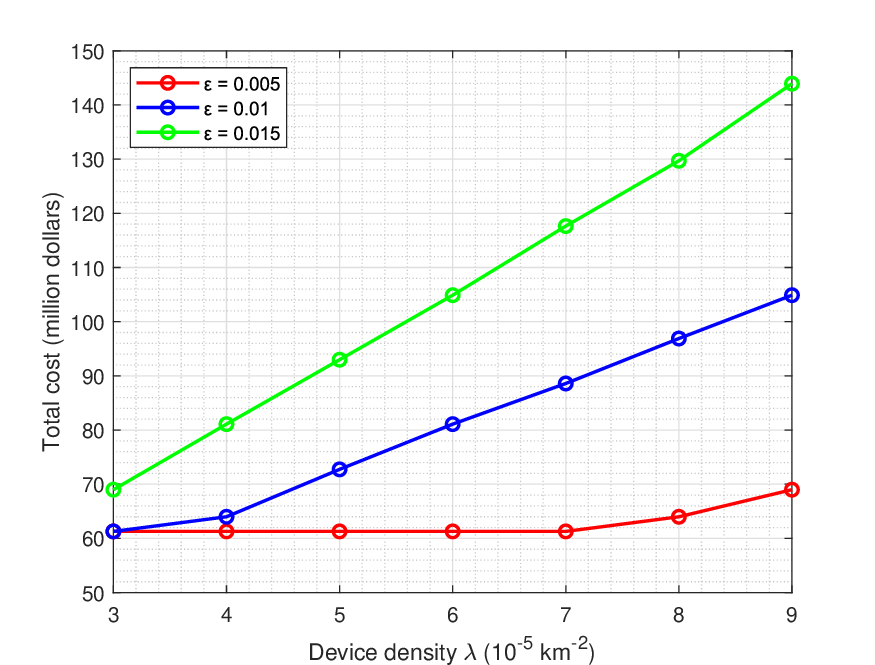}
        \caption {The total cost of LEO satellite IoT constellation under different device densities and activity probabilities $\varepsilon$.}
        \label{pa}
    \end{figure}

    LEO satellite IoT constellation needs to provide service for diverse IoT devices in different scenarios, e.g., ocean, desert, forests and mountains. Generally, these IoT devices may have quite different device densities and activity probabilities. For this purpose, we investigate the relation between the total cost of the LEO satellite constellation and the IoT device densities and activity probabilities. It can be seen in Fig. \ref{pa} that for a given activity probability $\varepsilon$, as the device density increases, the total cost increases. Moreover, the total cost is added as the activity probability increases. This is because more activated IoT devices need to be supported by LEO satellites in a certain area, resulting in the increment of the number of LEO satellites in a satellite constellation. Additionally, when the product of device density and activity factor is equal, the total cost of the designed constellation has the same value, {since the value of the activity factor times the device density is directly related to the average interference and backhaul capacity, as indicated in (19) and (20).}

    \begin{figure}
        \centering
        \includegraphics [width=0.45\textwidth] {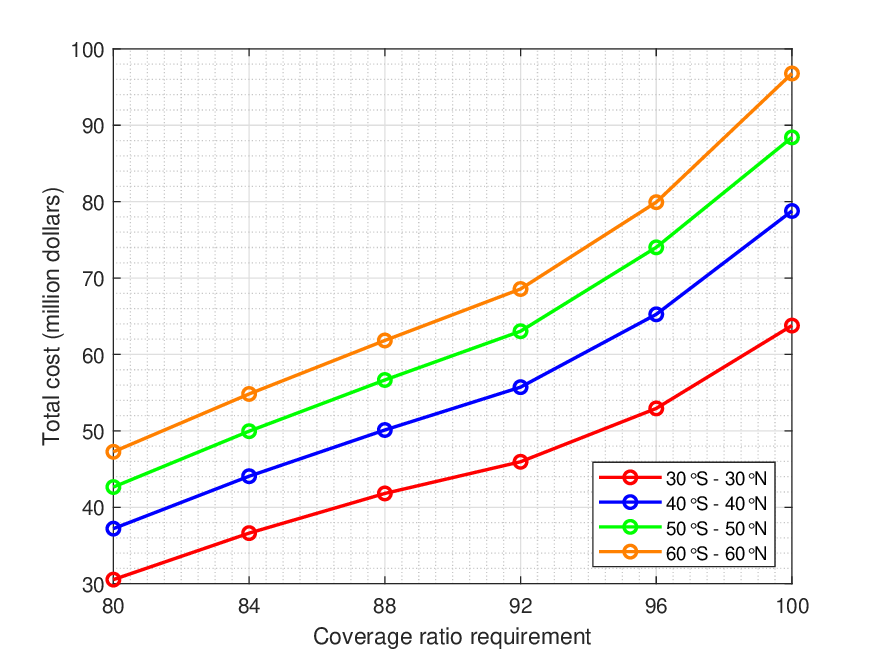}
        \caption { The total cost of LEO satellite IoT constellation with different coverage ratio and coverage area requirements. }
        \label{covlong}
    \end{figure}

    {Generally, the coverage performance is essential for the design of LEO satellite constellation. Therefore, the relationship between the total cost of the satellite constellation and the coverage ratio requirements has been shown in Fig. \ref{covlong} under different coverage areas. It can be found that $100\%$ coverage for the target area will result in a sharp increment in costs. Moreover, it is worth noticing that the proposed algorithm is applicable in the scenarios with different coverage areas. Specifically, for a given required coverage ratio, the cost is added as the coverage area is broadened.}

    \begin{figure}
        \centering
        \includegraphics [width=0.45\textwidth] {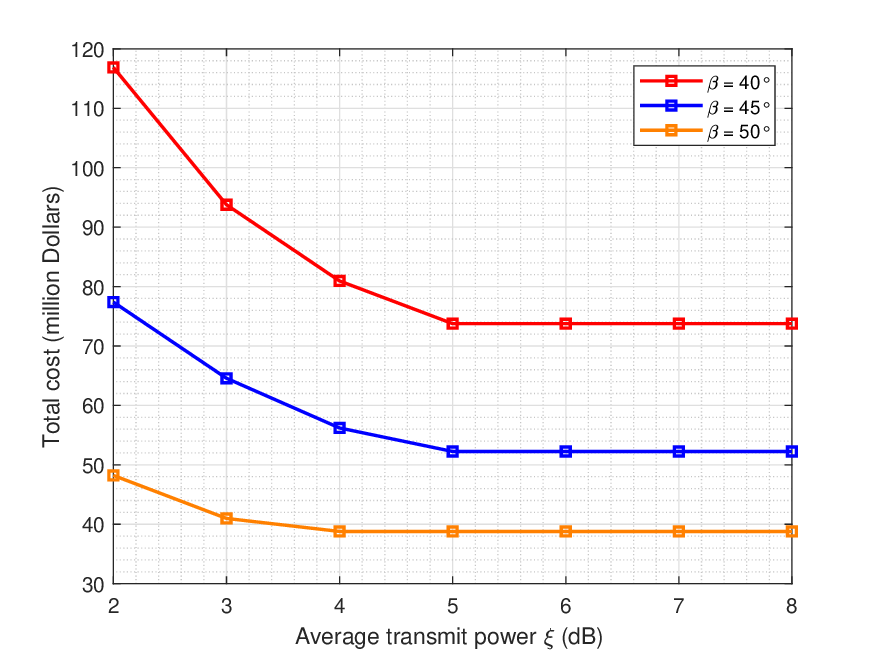}
        \caption { The total cost of LEO satellite IoT constellation under different satellite coverage angles $\beta$ and transmit power of devices $\xi$.}
        \label{pow}
    \end{figure}

    {For transmit power of IoT devices, it will also affect the performance of the LEO satellite IoT constellation.} Fig. \ref{pow} shows that increasing the transmit power of IoT devices can effectively reduce the total cost of the constellation. However, the total cost will come to a convergence as the transmit power of devices $\xi$ increases. Fortunately, as illustrated in Fig. \ref{pow}, a feasible way to reduce the total cost of LEO satellite IoT constellation is to equip the satellite with sensor antennas which has large coverage angles. In this way, designing sensor antennas with large coverage angles for LEO satellite is an important task in the future of satellite constellation.

    \begin{figure}
        \centering
        \includegraphics [width=0.45\textwidth] {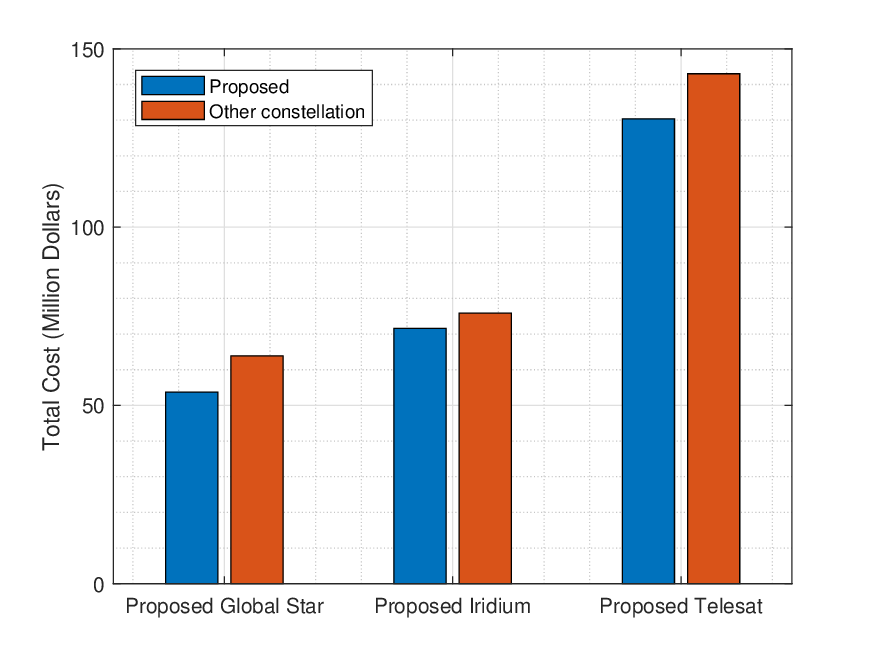}
        \caption { The total cost of the proposed LEO satellite constellation with different constellations.}
        \label{cost}
    \end{figure}

    Finally, we compare the proposed LEO satellite constellation with three classical LEO satellite constellations under identical communication backhaul capacity and coverage ratio requirements \cite{E1}-\cite{E3}. It can be seen from Fig. \ref{cost} that the proposed LEO satellite constellation is cost-competitive compared to other constellations, meeting our desired objectives. Furthermore, it is worth pointing out that the results in Fig. \ref{cost} do not mean that other constellations are useless. Actually, all the constellations are designed for different tasks. Specifically, Global Star is to provide seamless coverage for users except North and South Poles. Iridium aims to provide global communication service. Telesat wants to provide cost-effective, fiber-like, global network connectivity for business, government and individual users. Fortunately, the proposed algorithm can minimize the deployment cost while guaranteeing the coverage ratio and communication quality.

   \section{Conclusion and Future Work}
       In this paper, we provided a QoS-driven satellite constellation design scheme for LEO satellite IoT. In particular, a LEO satellite IoT constellation optimization algorithm was proposed to minimize the deployment cost while guaranteeing the coverage ratio and communication quality of IoT devices. Theoretical analysis proved that the proposed algorithm can achieve the global convergence. Finally, extensive simulation results validated the effectiveness of the proposed algorithm. Specifically, the proposed algorithm can converge in $10$ iterations and obtain a lower satellite constellation cost than the baseline optimization algorithms.

       {As for future work, the paper suggests several extensions. These include integrating dynamic traffic demand models, exploring multi-objective optimization with real-time adaptive algorithms, and investigating the impact of inter-satellite links on constellation performance. These extensions would provide valuable insights for next-generation LEO satellite constellation design.}

   \begin{appendices}
   	\section{The Proof of Theorem 1}
   Notice that the activation indicator $\alpha_k$, data sequence $\mathbf{s}_k$, large-scale fading factor $g_{m,k}$ and small-scale fading factor $\widetilde{\mathbf{h}}_{m,k}$ are mutually independent, such that the average interference $\mathbb{E}[I_{m,k}]$ can be decomposed into a product of the expectations of these variables as
   	\begin{equation}\label{einf1}
    \begin{aligned}
        \mathbb{E} [I_{m,k}] = &\xi\sum_{\begin{subarray}{c}k'\in\mathcal{D}_{m,k}\\k'\neq k\end{subarray}} \mathbb{E} [\alpha_k] \cdot \mathbb{E} [\mathbf{s}_{k'}^H\mathbf{s}_{k'}]\cdot g_{m,k'}^2 \cdot \mathbb{E} [\widetilde{\mathbf{h}}_{m,k'}^H\widetilde{\mathbf{h}}_{m,k'}] \\
        &= \xi\varepsilon L M_a \sum_{\begin{subarray}{c}k'\in\mathcal{D}_{m,k}\\k'\neq k\end{subarray}} g_{m,k'}^2,\\
    \end{aligned}
   	\end{equation}
    {where $\mathbf{s}_k$ denotes the transmitted data sequence of device $k$, $L$ and $M_a$ represents the length of data sequence and the number of antennas, respectively.} As the distribution of IoT devices follows HPPP with density $\lambda$, according to the Campbell’s theorem \cite{tho}, the average interference $\mathbb{E}[I_{m,k}]$ can be derived as
   	
   	\begin{equation}\label{einf2}
    \begin{aligned}
         \mathbb{E} [I_{m,k}]& = \xi\varepsilon L M_a\sum_{\begin{subarray}{c}k'\in\mathcal{D}_{m,k}\\k'\neq k\end{subarray}} g_{m,k'}^2 \\
        & = \xi\varepsilon L M_a \lambda\int_{k'\in\mathcal{D}_{m,k}}g_{m,k'}^2 dk' .\\
    \end{aligned}
   	\end{equation}
   The area of $\mathcal{D}_{m,k}$ is a spherical crown with the angular radius $\varphi_0$, which motivates us to use the spherical segments as the partition of the integral region, then the average interference $\mathbb{E} [I_{m,k}]$ can be obtained as
   	\begin{equation}\label{einf3}
   		\begin{aligned}
   			\mathbb{E} [I_{m,k}] &= \xi\varepsilon L M_a \lambda\int_{k'\in\mathcal{D}_{m,k}}g_{m,k'}^2 dk' \\
   			&= \xi\varepsilon L M_a \lambda (\frac{c}{4\pi f_c})^2\cdot G_{\text{sat}}G_{\text{dev}}\cdot \frac{1}{r_0} \int_{k'\in\mathcal{D}_{m,k}}d_{m,k'}^{-2} dk'\\
   			&= \xi\varepsilon L M_a \lambda (\frac{c}{4\pi f_c})^2\cdot G_{\text{sat}}G_{\text{dev}}\cdot \frac{1}{r_0} \int_{0}^{\text{arccos} \frac{R_e}{R_e+h}} [R_e^2\\
                &+(R_e+h)^2-2R_e(R_e+h)\text{cos}\varphi]^{-1} 2\pi R_e \text{sin}\varphi R_e d\varphi\\
   			&= \xi\varepsilon L M_a \lambda (\frac{c}{4\pi f_c})^2                             \cdot G_{\text{sat}}G_{\text{dev}}\cdot \frac{1}{r_0}
   			2\pi R_e^2 \int_{0}^{\text{arccos}\frac{R_e}{R_e+h}} \\
                &[R_e^2+(R_e+h)^2-2R_e(R_e+h)\text{cos}\varphi]^{-1}  d(-\text{cos}\varphi)\\
   			&=\xi\varepsilon L M_a \lambda (\frac{c}{4\pi f_c})^2\cdot G_{\text{sat}}G_{\text{dev}}\cdot \frac{1}{r_0}	2\pi R_e^2\ln(2\frac{R_e}{h}+1).
   		\end{aligned}
   	\end{equation}
     The proof is completed.
   	
   	\section{The Proof of Theorem 2}
     For the expectation of term $\log_2 (1+\Gamma_{m,k})$ in (\ref{SINR2}), it can be further expressed as
   	\begin{equation}\label{ega}
   		\begin{aligned}
   			\mathbb{E} [\log_2 (1+&\Gamma_{m,k})]  = \mathbb{E} [\log_2 (1+\frac{ \xi L M_a g_{m,k}^2}{\mathbb{E}[I_{m,k}]+\sigma_0^2})] \\
   			  &= \frac{1}{S_{\mathcal{A}_k}\cdot \ln 2}\int_{m\in\mathcal{A}_k}\ln \bigg (1+\\
        &\frac{\xi L M_a\cdot(\frac{c}{4\pi f})^2\cdot G_{\text{sat}}G_{\text{dev}}\cdot\frac{1}{r_0}}{\mathbb{E}[I_{m,k}]+\sigma_0^2}d_{m,k}^{-2} \bigg) dm.
   		\end{aligned}
   	\end{equation}
   where $\xi$ and $\varepsilon$ denote the average transmit power and average activity probability over all IoT devices, respectively, {$S_{\mathcal{A}_k}$ denotes the area of visible satellite region of device $k$.} To make the expression more clearly, we use $\Psi$ to denote the coefficient of the integral variable, i.e., $\Psi = \frac{\xi L M_a\cdot(\frac{c}{4\pi f_c})^2\cdot G_{\text{sat}}G_{\text{dev}}\cdot\frac{1}{r_0}}{\mathbb{E}[I_{m,k}]+\sigma_0^2}$. Then, the expectation can be rewritten as
   	\begin{equation}\label{ega}
   		\begin{aligned}
   			\mathbb{E} &[\log_2 (1+\Gamma_{m,k})]  = \frac{1}{S_{\mathcal{A}_k}\cdot \ln 2}\int_{0}^\varphi\ln (1+ \Psi\cdot [R_e^2+\\
      &(R_e+h)^2-2 R_e(R_e+h) \cos v]^{-1} ) 2\pi(R_e+h)^2 \mathrm{sin} v dv. \\
   			& \overset{(a)}{=} \frac{\pi (R_e+h)}{S_{\mathcal{A}_k}\ln 2\cdot R_e}\int_{h^2}^{d_m^2} \ln (1+ \Psi u^{-1}) du,\\
   			& = \frac{\pi (R_e+h)}{S_{\mathcal{A}_k}\ln 2\cdot R_e} \Big[\ln(1+\Psi d_m^{-2})d_m^{2}\\
                &+\Psi \ln(d_m^{2}+\Psi)-\ln(1+\Psi h^{-2})d_m^{2}-\Psi \ln(h^{2}+\Psi) \Big]
   		\end{aligned}
   	\end{equation}
   	where in (a) the substitution  $u = R_e^2+(R_e+h)^2-2R_e(R_e+h)\cos v$ is applied.

    The proof is completed.

    \section{The Proof of the global convergence of proposed algorithm}
     For the sake of convenience in the proof, we present the following important concepts.

     \emph{Definition 3}: Let $\{X_k:k\geq 1\}$ denote a sequence of random variables defined on a sample space $\Omega$. We say that $\{X_k:k\geq 1\}$ is convergent in probability to a random variable $X$ if and only if
     \begin{equation}\label{def4}
       %  \lim_{k\rightarrow\infty}\mathbb{P}(|X_k-%X|\leq\varepsilon)=1,
        \mathbb{P}(\bigcap_{t=1}^{\infty}\bigcup_{k\geq t}||X_k-X||\geq \delta)=0,
     \end{equation}
     for any $\delta>0$.

     \emph{Lemma 1 (Borel–Cantelli lemma, BCL)}: Suppose $\{A_k:k\geq 1\}$ denotes a sequence of random variables defined on a sample space $\Omega$. For simplicity, we let $p_k = \mathbb{P}(A_k)$, if $\sum_{k=1}^{\infty}p_k <\infty$, then
     \begin{equation}\label{lem1}
         \mathbb{P}(\bigcap_{t=1}^{\infty}\bigcup_{k\geq t}A_k)=0,
     \end{equation}
    else if $\sum_{k=1}^{\infty}p_k =\infty$ and $\{A_k:k\geq 1\}$ are mutual independent, then
    \begin{equation}\label{lem1}
         \mathbb{P}(\bigcap_{t=1}^{\infty}\bigcup_{k\geq t}A_k)=1.
     \end{equation}
     The detailed proof of Borel–Cantelli lemma can be found in \cite{BCL}.

     For $\forall \delta>0$, we define $Q_1$ which satisfies
    \begin{equation}\label{Q1}
     \mathcal{B}_1= \{x\in D:|f(x)-f_{\min}(x)|<\delta\},
     \end{equation}
     where $D$ denotes the feasible domain of the optimization problem, and $f_{\min}(x)=\mathrm{\min}\{f(x),x\in D\}$. Moreover, we define $B_2$ as the complementary set of $B_1$, i.e.,
    \begin{equation}\label{Q2}
     \mathcal{B}_2= \complement_D \mathcal{B}_1,
     \end{equation}
    In this way, the population $\{P(t)\}$ processed by the proposed algorithm can be divided two states: $\mathcal{S}_1$ and $\mathcal{S}_2$. If there exists one individual of $\{P(t)\}$ belonging to $B_1$, then it means that the population $P(t)$ is in state $\mathcal{S}_1$, i.e., $P(t)\in\mathcal{S}_1$. Else if all the individuals of $P(t)$ belong to $\mathcal{B}_2$, then it means that the population $\{P(t)\}$ is in state $\mathcal{S}_2$, i.e., $P(t)\in\mathcal{S}_2$. Based on it, we can derive the following lemma:

    \emph{Lemma 2}: Let $p_{ij}$ be the probability of $P(t)$ in $\mathcal{S}_i (i=1,2)$ while
   $P(t+1)$ in $\mathcal{S}_j (j=1,2)$. Then we have:
   \begin{enumerate}
   \item For $P(t)\in \mathcal{S}_1$, we have $p_{11} =1.$
    \item For $P(t)\in \mathcal{S}_2$, there exists $c\in(0,1)$ which satisfies $p_{22}\leq c$.
   \end{enumerate}
    \begin{IEEEproof}[Proof]
	In the proposed algorithm, we adopt elitist selection as the choice method for populations, if $P(k)\in \mathcal{S}_1$, then $P(k)\in \mathcal{S}_1$, i.e., $p_{11} =1$. Therefore, 1) is proved.

    To prove 2), we assume that $x^*$ is the optimal solution. Then $\exists \gamma>0$, there exists
    \begin{equation}\label{xstar}
        |f(x)-f(x^*)|<\frac{\delta}{2},
    \end{equation}
    for $\forall x\in D \bigcap N_{\gamma}(x^*)$, where $N_{\gamma}(x^*)=\{x\in S:|x-x^*|\leq\gamma\}$. Based on (\ref{xstar}), we can obtain
    \begin{equation}\label{Q1}
        D \cap N_{\gamma}(x^*)\subseteq \mathcal{B}_1.
    \end{equation}
    In the case of $P(t)\in \mathcal{S}_2$, we assume $\mathbf{x}_{mu}$ is the results of the proposed algorithm after step 4 with population $P(k)$. According to (\ref{xmu1}) and (\ref{xmu2}), the probability of $\mathbf{x}_{mu}=\mathbf{x}+\Delta \mathbf{m}$ can be expressed as
    \begin{equation}\label{p1}
        \mathbb{P}(\mathbf{x}_{mu}=\mathbf{x}+\Delta \mathbf{m}) = (\frac{1}{1+t}+1)\varsigma=p_1.
    \end{equation}
    Based on (\ref{p1}), the probability of
    $\mathbf{x}_{mu}\in N_{\gamma}(x^*)\bigcap D$ can be calculated as
    \begin{equation}\label{xchi}
    \begin{aligned}
        \mathbb{P}(\mathbf{x}_{mu}&\in N_{\gamma}(x^*)\cap D)=p_1*\mathbb{P}(\mathbf{x}+\Delta \mathbf{m}\in N_{\gamma}(x^*)\cap D)\\
       & \geq (\frac{1}{1+t}+1)\varsigma\cdot\prod_{i=1}^n\mathbb{P}(|x_i+\Delta m_i-x^*_i|\leq \gamma)\\
       &= (\frac{1}{1+t}+1)\varsigma\cdot\prod_{i=1}^n\int_{x_i^*-x_i-\gamma}^{x_i^*-x_i+\gamma}\frac{1}{\sqrt{2\pi}\sigma_i}e^{-t^2/2\sigma_i^2} dt\\
       & \geq\varsigma\prod_{i=1}^n\int_{x_i^*-x_i-\gamma}^{x_i^*-x_i+\gamma}\frac{1}{\sqrt{2\pi}\sigma_i}e^{-t^2/2\sigma_i^2} dt
    \end{aligned}.
    \end{equation}
    where $x_i$ and $x_i^*$ denote the $i$-th element of $\mathbf{x}$ and $\mathbf{x}^*$, respectively. For clarity, we use $p^*(\mathbf{x})$ to denote $\mathbb{P}(\mathbf{x}+\Delta \mathbf{m}\in N_{\gamma}(x^*)\cap D)$. As $ N_{\gamma}(x^*)\cap D$ is a bounded, non-void, closed set, and based on (\ref{xchi}), we can obtain
    \begin{equation}\label{xchi}
         0<p^*(\mathbf{x})<1.
    \end{equation}
    Moreover, as $D$ is a bounded closed set, we can obtain
    \begin{equation}\label{xchi}
         \exists \hat{\mathbf{x}}\in D, P^*(\hat{\mathbf{x}}) = \min\{p^*(\mathbf{x})|\mathbf{x}\in D\}.
    \end{equation}
    Based on  (\ref{Q1}) and (\ref{xchi}), we can get
    \begin{equation}\label{xchi2}
     0<p^*(\hat{\mathbf{x}})\leq p^*(\mathbf{x}) \leq p_{21}.
    \end{equation}
    Supposing $c=1-p^*(\hat{\mathbf{x}})$, we can obtain
    \begin{equation}\label{c}
     c = 1-p^*(\hat{\mathbf{x}})\geq 1-p_{21}=p_{22}.
    \end{equation}
    The Lemma 2.2) is proved.
   \end{IEEEproof}

    Based on Lemma 2, we denote $p_t = \mathbb{P}(|f(\mathbf{x}_t^*)-f(\mathbf{x}^*)|\geq \delta)$, where $\mathbf{x}_t^*= \min_{P(t)\cap D} f(\mathbf{x})$, then we can get
    \begin{equation}\label{pt}
        p_t = \left\{
        \begin{aligned}
            &0 \quad \quad \exists i, \mathbf{x}_i^*\in \mathcal{B}_1\\
            &\hat{p}_t \quad\quad \forall i,   \mathbf{x}_i^* \notin \mathcal{B}_1
        \end{aligned}
        \right..
    \end{equation}
    According to Lemma 2, $\hat{p}_t$ can be estimated by $c$ as
    \begin{equation}\label{pt2}
         \hat{p}_t = \mathbb{P}(\mathbf{x}_t^*\notin Q_1, i=1, 2, \cdots, t)=p_{22}^t\leq c^t.
    \end{equation}
    Therefore, we can obtain
    \begin{equation}\label{sumpt}
         \sum_{t=1}^{\infty}p_t\leq \sum_{t=1}^{\infty}c^t =\frac{c}{1-c} <\infty.
    \end{equation}
    According to Lemma 1, we can obtain
    \begin{equation}\label{123}
        \mathbb{P}(|f(\mathbf{x}_t^*)-f(\mathbf{x}^*)|\geq \delta )=0.
    \end{equation}
   Finally, based on \emph{Definition 3}, it is proved that the proposed algorithm achieves global convergence.

   The proof is completed.
\end{appendices}

\end{document}